\documentclass[seceq]{ptptex}

\usepackage{graphicx}
\usepackage{wrapft}

\newcommand{\e}{{\rm e}}

\newcommand{\EP}{\mathcal{E}}

\def\eqalign#1{\,\vcenter{\openup1\jot\ialign
    {\strut\hfil$\displaystyle{##}$&$\displaystyle{{}##}$
      \hfil\crcr#1\crcr}}\,}

\def\caseNA{{[{\rm n}_{\rm A}]}}
\def\caseNB{{[{\rm n}_{\rm B}]}}
\def\caseI{{[~{\rm i}~]}}
\def\caseII{{[~{\rm ii}~]}}

\title{Light propagation and gravitational lensing\\
on the Weyl-like spacetime\\
in scalar-tensor theories of gravity
}

\author{
Shin-ei \textsc{Tsuneishi}$^{1}$
\thanks{
tuneishi@astro.sc.niigata-u.ac.jp}\
,
~Kazuya \textsc{Watanabe}$^{2}$
\thanks{
kazuya@astro.sc.niigata-u.ac.jp}\\
and
~Tooru \textsc{Tsuchida}$^{3}$
\thanks{
CZL03253@nifty.ne.jp}\\
}

\inst{
$^1$ Graduate School of Science and Technology, Niigata University
  Niigata, 950--2181, Japan\\
$^2$ Department of Physics, Niigata University,
         Niigata 950--2181, Japan\\
$^3 $Kokugakuin~University~Tochigi~High~School,~Tochigi~328-8588,~Japan.
}

\abst{

\

We study light propagation and gravitational
lensing in scalar-tensor theories of gravity
by using a static, axisymmetric exterior solution.
The solution, which has been studied by Tsuchida and Watanabe,
has asymptotic flatness properties 
and moreover 
is reduced to Voorhees's one of Weyl's series of prolate solutions
in the case of a constant scalar field.
We first obtain an asymptotic form of the solution
near the spatial infinity in order to clarify the physical
significance of three model parameters found in the solution.
It is shown that 
an amplitude of the scalar field, 
a non-sphericality of the spacetime symmetry and
a mass-like parameter in the Einstein frame
are represented in terms of these three parameters. 
As for the spacetime structure, we also give brief discussion on
the directional-dependent properties of the spacetime singularity.
We then study null geodesic equations 
and Sachs's optical scalar equations on the equatorial plane
in order to investigate deflection and shear of light rays.
Our studies are done by using a technique of
the conformal transformation such that their results
are independent of details of scalar-tensor theories of gravity
owing to the conformally invariant properties of null geodesics.
For some specific values of the model parameters, 
we analytically obtain a deflection angle of the light path 
and find that it can become negative.
The appearance of a negative deflection angle indicates 
``reflection'' of a light path, and we investigate under
which conditions the light reflection occurs. 
As for the optical scalars, we find that
the Weyl source-term shows significantly different properties
when it is compared with that in the Schwarzschild spacetime.
We therefore classify a space of the model parameters
into four distinct regions on the basis of the qualitative
properties of the Weyl source-term
and find a close relationship between this classification 
of the parameter space and the occurrence of the light reflection. 
We finally solve the null geodesic equations 
and the optical scalar equations numerically.
We find that a picture of the thin lens is applicable
and give a simple analytic model for the optical scalars.
As for the properties of gravitational lensing,
the deflection angle and the image distortion rate 
are obtained as functions of
the impact parameter. Again, we find a close relationship 
between their qualitative properties and the classification above.

}

\begin{document}

\maketitle

\
\

\section{Introduction}

\

Scalar-tensor theories of gravity
have been studied by many theoretical
physicists as natural alternatives to general relativity
since the pioneering work by Brans and Dicke\cite{1}.
In particular, a theory of dilaton gravity 
has been of interest as an effective theory of the superstring theories 
at low energy scales\cite{2}.
In these theories, the gravity is mediated not only by a
tensor field but also by a scalar field.
Several theoretical predictions in the scalar-tensor theories
have been obtained$,^{3)-7)}$ and 
it has been found that a wide class of the scalar-tensor theories can pass 
all the experimental tests in the cases of weak gravitational fields.
In the cases of strong gravitational fields, however, 
it has been found that the scalar-tensor theories
show different aspects of the gravity in contrast 
to general relativity.
For example, it has been shown numerically 
that nonperturbative effects in the scalar-tensor 
theories increase the maximum mass of an isolated system
such as a neutron star$.^{3)-5)}$
In these works, numerical methods play an important role,
and a spherical exterior solution, which is 
a static, spherically symmetric exterior solution
outside a spherical body, is matched to
the numerical solution in interpreting its results.
A usual astronomical object, however, 
has a non-spherical form and is in rotation, and 
not only a spherical exterior solution but also 
axisymmetric exterior solutions may therefore play an important role
in discussing the crucial difference between general relativity
and scalar-tensor theories of gravity.

Among the axisymmetric exterior solutions, 
ones corresponding to the series of solutions 
of Tomimatsu and Sato in general relativity\cite{8}
must be of significant interest.
It is, however, difficult
to obtain their explicit forms, and the case is similar
even for solutions corresponding to
Kerr's solution in general relativity. 
Instead of studying such the Kerr-like solutions, 
Tsuchida and Watanabe\cite{7} have investigated
the Weyl-like solutions, namely 
the solutions which are reduced to those of Weyl's series 
in the case of a constant scalar field. 
Among these Weyl-like solutions, we have a special solution, 
which we shall refer to as
the scalar-tensor-Weyl solution.
This solution is very notable for the reason
that it has asymptotic flatness
properties and moreover is reduced to Voorhees's solution\cite{9}, 
namely a solution of Weyl's series of prolate solutions,
in the case of a constant scalar field.
Several geometrical properties of
the scalar-tensor-Weyl solution, especially the properties
of light propagation, have been investigated by
analytical approaches in Ref.7. On the basis of 
this work, we will study light propagation and gravitational lensing
on the scalar-tensor-Weyl solution by numerical approaches
because the local properties of null geodesics
must tell us important geometrical information of the spacetime.

In this paper, we first study the global structure of
the scalar-tensor-Weyl solution.
We obtain an asymptotic form of the solution
near the spatial infinity in order to clarify the physical
significance of three model parameters found in the solution.
It may be interesting that the Schwarzschild-like coordinates
can be naturally introduced in doing that. 
It will be shown that these parameters are related to 
an amplitude of the scalar field,
a non-sphericality of the spacetime symmetry and 
a mass-like parameter in the Einstein frame.
The directional-dependent properties of the spacetime singularity
will be also shown.
After giving the several analytical results as briefly
summarized in Abstract,
we numerically solve null geodesic equations and Sachs's optical 
scalar equations\cite{10}.
By using a technique of
the conformal transformation, 
we then study deflection and shear
of light rays on the equatorial plane 
in the manner such that the obtained results are
independent of details of scalar-tensor theories of gravity.
We numerically obtain the deflection angle and
the image distortion rate as functions of
the impact parameter and find a close relationship 
between their qualitative properties and the classification 
of the model parameter space according to
the Weyl source-term.
Several analytic results shown in the previous papers are
summarized in Appendices\cite{7,11}.

Throughout this paper, we use the system of units such that $c=G=1$.

\

\section{The scalar-tensor-theory of gravity}

\

We consider the simplest scalar-tensor theory
in which gravitational
interactions are mediated by a tensor field, $ \hat{g}_{\mu\nu} $, and
a scalar field, $ \hat{\phi}.^{1),3)-7)}$  
Hereafter, a hat (~$\hat{~}$~) is used to denote
quantities and derivatives associated with $\hat{g}_{\mu\nu}$.  An
action of the theory is given as
\begin{equation}
  S=\frac{1}{16\pi}\int
    \left[\hat{\phi}
    \hat{R}-\frac{\omega(\hat{\phi})}
    {\hat{\phi}}\hat{g}^{\mu\nu}\hat{\phi}_{,\mu}\hat{\phi}_{,\nu}
    \right]\sqrt{-\hat{g}}d^4x +
  S_{\mbox{\tiny matter}}[\hat{\Psi}_{\mbox{\tiny
      m}},\hat{g}_{\mu\nu}],~~\hat{\phi}_{,\mu}\equiv
          \frac{\partial\hat{\phi}}{\partial x^\mu},   
\label{ea5}
\end{equation}
where $\omega(\hat{\phi})$ is a dimensionless 
function of $\hat{\phi}$ and specifies the theory,
$\hat{\Psi}_{\mbox{\tiny m}}$ represents matter fields, and
$S_{\mbox{\tiny matter}}$ is an action of the matter fields.  The
scalar field, $\hat{\phi}$, plays the role of an effective gravitational
constant, $\hat{G}$, as $\hat{G}\sim1/\hat{\phi}$.  
Varying the action by the tensor
field, $\hat{g}_{\mu\nu}$, and the scalar field, $\hat{\phi}$, yields,
respectively, the following field equations,
\begin{eqnarray}
  \hat{G}_{\mu\nu}& = & 
     \frac{8\pi}{\hat{\phi}}\hat{T}_{\mu\nu}
     +\frac{\omega(\hat{\phi})}{\hat{\phi}^2}
     \left(\hat{\phi}_{,\mu}\hat{\phi}_{,\nu}-\frac{1}{2}\hat{g}_{\mu\nu}
      \hat{g}^{\alpha\beta}\hat{\phi}_{,\alpha}
      \hat{\phi}_{,\beta}\right)\nonumber\\
    &~& +
      \frac{1}{\hat{\phi}}(\hat{\nabla}_{\mu}\hat{\phi}_{,\nu}
      -\hat{g}_{\mu\nu} 
     \hat{\nabla}^\alpha\hat{\phi}_{,\alpha}),\label{ea6} \\ 
\hat{\nabla}^\alpha
\hat{\phi}_{,\alpha} & =
  & \frac{1}{3+2\omega(\hat{\phi})}\left(8\pi \hat{T} -
    \frac{d\omega(\hat{\phi})}{d\hat{\phi}}
    \hat{g}^{\alpha\beta}\hat{\phi}_{,\alpha}
    \hat{\phi}_{,\beta}\right),
\label{ea7}
\end{eqnarray}
where $\hat{\nabla}_\alpha$ and $\hat{T}_{\mu\nu}$ 
denote, respectively, a covariant 
derivative and an energy-momentum tensor
associated with $\hat{g}_{\mu\nu}$. 
Now we perform the following conformal transformation
to a new frame called the Einstein frame with the metric,
$g_{\mu\nu}$, defined by
\begin{equation}
  g_{\mu\nu}=A^{-2}(\varphi)\hat{g}_{\mu\nu},
\label{ea8}
\end{equation}
such that
\begin{equation}
  A^2(\varphi) = \frac{1}{\hat{\phi}},~~
 \frac{1}{3+2\omega(\hat{\phi})} 
   =\left(\frac{d\ln A(\varphi)}{d\varphi}\right)^2 
   \equiv\alpha^2(\varphi),
\label{ea11}
\end{equation}
where $A(\varphi)$ and $\alpha(\varphi)$ 
are referred to as a coupling function and a coupling strength,
respectively. Then the action is rewritten as
\begin{equation}
  S=\frac{1}{16\pi
    }\int
  (R-2g^{\mu\nu}\varphi_{,\mu}\varphi_{,\nu})\sqrt{-g}d^4x +
  S_{\mbox{\tiny matter}}[\hat{\Psi}_{\mbox{\tiny m}},A^2(\varphi)
  g_{\mu\nu}].
\label{ea10}
\end{equation}
Varying the action by $g_{\mu\nu}$ and $\varphi$ yields,
respectively,
\begin{eqnarray}
  G_{\mu\nu}& = & 8\pi T_{\mu\nu} +
  2\left(\varphi_{,\mu}\varphi_{,\nu}
    -\frac{1}{2}g_{\mu\nu}g^{\alpha\beta}\varphi_{,\alpha}
    \varphi_{,\beta}\right),
\label{ea12} \\
\nabla^\mu\varphi_{,\mu}& = &-4\pi\alpha(\varphi) T,
\label{fs2}
\end{eqnarray}
where $\nabla_\alpha$ and $T_{\mu\nu}$ denote, respectively, 
a covariant derivative and an energy-momentum tensor
associated with $g_{\mu\nu}$. 
A relationship between ${T}^{\mu\nu}$
and $\hat{T}^{\mu\nu}$ is given by
\begin{equation}
 T^{\mu\nu} \equiv \frac{2}{\sqrt{-g}} \frac{\delta S_{\mbox{\tiny
        matter}}[\hat{\Psi}_{\mbox{\tiny m}},A^2(\varphi)g_{\mu\nu}]}
  {\delta g_{\mu\nu}}
=A^6(\varphi)\hat{T}^{\mu\nu}.
\label{ea14}
\end{equation}
The conservation law for $T^{\mu\nu}$ is given by
\begin{equation}
  \nabla_{\nu}T^{\mu\nu}=\alpha(\varphi)T \nabla^{\mu}\varphi.
\label{ea15}
\end{equation}

\

\section{Geometrical properties of the scalar-tensor-Weyl solution}

\

We investigate the geometrical properties of 
the scalar-tensor-Weyl solution\cite{7}.
The reduced field equations and the derivation of the solution
are given in Appendices A and B.

\

\subsection{The scalar-tensor-Weyl solution}

\

The scalar-tensor-Weyl solution is 
one of the static, axisymmetric vacuum solutions
in the Einstein frame with the metric,
\begin{eqnarray}
\eqalign{
ds^2 & =  -\left(\frac{x-1}{x+1}\right)^{\delta} dt^2 
+\sigma^2 \left(\frac{x-1}{x+1}\right)^{-\delta} \times\cr
& 
\left[\left(\frac{x^2-1}{x^2-y^2}\right)^{\Delta^2}
(x^2-y^2)\left(\frac{dx^2}{x^2-1}+\frac{dy^2}{1-y^2}\right)+(x^2-1)(1-y^2)
d\phi^2\right],\cr}
\label{st1}
\end{eqnarray}
and the scalar field,
\begin{equation}
\varphi = \varphi_0 + \frac{d}{2}\ln\left(\frac{x-1}{x+1}\right) ,
\label{st2}
\end{equation}
where a positive constant, $\sigma$, is a unit of length,  
and $\delta$, $\varphi_0$ and $d$ are integration constants.
A parameter, $\Delta$, is defined by
\begin{equation}
{\Delta}^2\equiv \delta^2 +d^2,\ \ \Delta\geq0.\label{st3}
\end{equation}

Now we show that the above solution is reduced to
the previously known ones when we take specific values of 
the parameters, $\delta$ and $\Delta$. 
When $\Delta=\delta$, i.e., $d=0$, 
the metric is reduced to Voorhees's prolate solution\cite{9}
which is one of Weyl's series of the solutions
in general relativity. 
Voorhees's prolate solution contains the Schwarzschild solution
as a specific case, $\delta=1$, and the  
Schwarzschild coordinates are related to the coordinates, $x$ and $y$, as
\begin{equation}
x = \frac{r}{\sigma}-1, \ \ y=\cos\theta, \ \ m=\sigma,
\label{st4}
\end{equation}
where $m$ plays the role 
of a usual mass parameter of the Schwarzschild spacetime.

When $\Delta=1$, we introduce the Just coordinates by
\begin{equation}
x = \frac{\chi}{\sigma}-1, \ \
y = \cos\theta, \ \
m=\delta \sigma.
\label{st5}
\end{equation}
Then (\ref{st1}) is reduced to a spherical exterior solution,
\begin{equation}
\eqalign{
ds^2 = &-\left(1-\frac{2m}{\delta\chi}\right)^{\delta}dt^2
+\left(1-\frac{2m}{\delta\chi}\right)^{-\delta}d\chi^2\cr
       &+\chi^2\left(1-\frac{2m}{\delta\chi}\right)^{1-\delta}
       \left(d\theta^2+\sin^2\theta d\phi^2\right).\cr}
\label{kyu}
\end{equation}
This result tells us that
the parameter, $\Delta$, has the physical significance of
a non-sphericality of the spacetime symmetry.
The Schwarzschild coordinate, $r$, is related to 
the Just coordinate, $\chi$, as
\begin{equation}
r=\chi \left(1-\frac{2m}{\delta\chi}\right)^{\frac{1-\delta}{2}}.
\label{kyur}
\end{equation}
When $\Delta=\delta=1$, 
the spherical exterior solution (\ref{kyu}) is reduced to
Schwarzschild's one, and, in this case, the scalar field 
becomes a constant.
A relationship among these
solutions is shown in Figure 1.

\begin{figure}
\begin{center}
\centerline{\includegraphics[width=.8\textwidth]{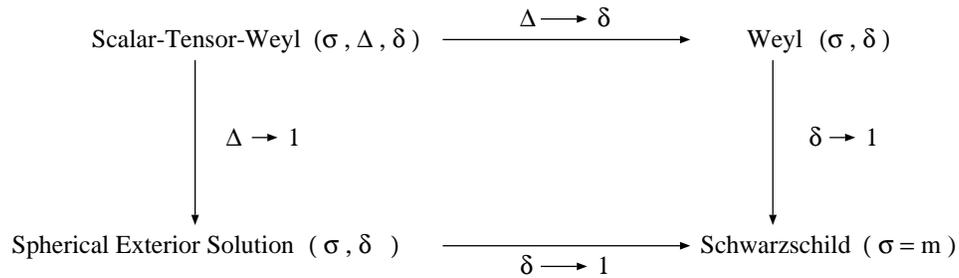}} 
\caption{A schematic depiction of series of 
the scalar-tensor-Weyl solutions.}
\label{fig.1}
\end{center}
\end{figure}

\

\subsection{An asymptotic form of the scalar-tensor-Weyl solution
near the spatial infinity}

\

Now, in order to clarify the physical significance of
the parameters, $\Delta$ and $\delta$,
we study an asymptotic form of the scalar-tensor-Weyl solution 
near the spatial infinity.
We first define the asymptotically Schwarzschild-like 
coordinates, $(r,\theta)$, by 
\begin{eqnarray}
x &=& \frac{\Delta}{\sqrt{2}}\left[u^2+v^2+\frac{1-\Delta^2}{\Delta^2}
+\sqrt{ 
\left( u^2+v^2 +\frac{1-\Delta^2}{\Delta^2}\right)^2
-\frac{4u^2 v^2}{\Delta^2} }
\right]^{ \frac{1}{2}}, 
\label{x}
\\
y &=&\pm\frac{\Delta}{\sqrt{2}}\left[u^2+v^2+\frac{1-\Delta^2}{\Delta^2}
-\sqrt{ 
\left( u^2+v^2 +\frac{1-\Delta^2}{\Delta^2}\right)^2
-\frac{4u^2 v^2}{\Delta^2} }
\right]^{ \frac{1}{2}}
\label{y},
\\
u &=& \frac{\delta}{m\Delta }\chi-1, \ \ 
v  =  \cos\theta, \ \ m=\delta \sigma,
\label{u}\\
\frac{m}{\chi} &=&
 \frac{m}{r}+\left( 1- \frac{\Delta}{\delta} \right)
\left\{\frac{m^2}{r^2}+
\left(3-\frac{\Delta}{\delta} 
\right)\frac{m^3}{2r^3}\right\}, 
\label{chi}
\end{eqnarray}
which are reduced to (\ref{st5}) when $\Delta=1$.
The equations in the above complicated forms are obtained 
under the following requirements.

\

\begin{description}
\item[(a)]~
The equations (\ref{x}) and (\ref{y}) are
the same as those in Voorhees's prolate solution,
and one has $\chi=r$ for $\Delta=\delta$ 
such that the asymptotic form of the solution
becomes that of Voorhees's one\cite{9}.
In particular, an asymptotic form 
of $g_{tt}$ for $\Delta=\delta$ becomes
\begin{eqnarray}
g_{tt} &=& -1+\frac{2m}{r}-\frac{2}{3}
\left(\frac{\delta^2-1}{\delta^2}\right)
\left(\frac{3}{2} \cos^2 \theta -\frac{1}{2}  \right) 
\frac{m^3}{r^3} + {\cal O}\left(\frac{m^4}{r^4}\right).
\label{weylgh} 
\end{eqnarray}

\item[(b)]~
When $\Delta=1$, 
the asymptotic form of the solution is the same 
as that of the spherical exterior solution, and
an asymptotic form of $g_{tt}$ 
and the relation between the Schwarzschild coordinate, $r$, and
the Just coordinate, $\chi$, are, respectively, given by
\begin{eqnarray}
g_{tt} &=& -1+\frac{2m}{r}+\frac{1}{3} \left(1-\frac{1}{\delta^2}
\right)\frac{m^3}{r^3}
+{\cal O}\left(\frac{m^4}{r^4}\right),  \label{kyugh} \\
\frac{m}{\chi} & = & \frac{m}{r}
+\left( 1- \frac{1}{\delta} \right)
\left\{
   \frac{m^2}{r^2}
   +\left(3-\frac{1}{\delta} \right)\frac{m^3}{2r^3}
   +{\cal O}\left(\frac{m^4}{r^4}\right)
\right\}. 
\label{kyuchih}
\end{eqnarray}

\item[(c)]~
For any values of $\Delta$ and $\delta$, 
asymptotic forms of $g_{tt}$ and $g_{\theta\theta}$ 
become, respectively, 
\begin{eqnarray}
g_{\theta\theta}
&=&r^2\left\{1+0\cdot\left(\frac{m}{r}\right)
+0\cdot\left(\frac{m^2}{r^2}\right)+
{\cal O}\left(\frac{m^3}{r^3}\right)\right\},\label{gtheta}\\
g_{tt}&=&-1+\frac{2m}{r}+0\cdot\left(\frac{m^2}{r^2}\right)
+{\cal O}\left(\frac{m^3}{r^3}\right).\label{gtt}
\end{eqnarray}
\end{description}

\

With (\ref{x})$\sim$(\ref{chi}), 
asymptotic forms of the metric and the scalar field 
near the spatial infinity are, respectively, given by
\begin{equation} 
\eqalign{
ds^2&=-\left[
1-\frac{2m}{r}+\left\{\frac{1}{3} 
\left(\frac{ \Delta^2}{\delta^2}-1
\right)
+\frac{2}{3} \left(\frac{\Delta^2-1}{\delta^2}\right)
\left(\frac{3}{2} \cos^2 \theta -\frac{1}{2}  \right)\right\} \frac{m^3}{r^3} 
\right]dt^2\cr
&~~+\left[
1-\frac{2m}{r}+\left(\frac{\Delta^2}{\delta^2}-1\right)
\frac{m^2}{r^2}
\right]^{-1}dr^2
+r^2\left(d\theta^2+\sin^2\theta d\phi^2\right),\label{new6}\cr
} 
\end{equation}
and
\begin{equation}
\eqalign{
\frac{\delta}{d} \left( \varphi -\varphi_0 \right) = 
&-\frac{m}{r}-\frac{m^2}{r^2}\cr
&-\left\{\frac{3}{2}
-\frac{1}{6} \left( \frac{\Delta}{\delta}\right)^2
-\left(\frac{\Delta^2-1}{3\delta^2}\right)\left(\frac{3}{2}\cos^2\theta-
\frac{1}{2} \right) \right\}\frac{m^3}{r^3}.\cr}
\label{scinf}
\end{equation}
Since ${\displaystyle
(d/\delta)^2=\Delta^2/\delta^2-1}$,
these asymptotic forms include
the parameters, $\Delta$ and $\delta$, 
in the forms, $\Delta^2/\delta^2-1$ and $(\Delta^2-1)/\delta^2$,
which can be interpreted as a squared effective amplitude 
of the scalar field
and a non-sphericality of the spacetime symmetry, respectively.
It is noted that the non-sphericality 
appearing in $g_{tt}$ can be read as a quadrupole moment of 
the gravitational potential in Newton's theory. 

\

\subsection{Directional properties of the spacetime singularity}

\

The scalar-tensor-Weyl solution has 
a coordinate singularity at $x=1$, and it is generally
a naked spacetime singularity
whose qualitative features are similar to
those of Voorhees's prolate solution.
A spatial topology of the singularity at $x=1$
is a sphere for $\Delta^2\neq1$ and becomes a point
for $\Delta^2=1$. 
When $\Delta^2=\delta=1$, the coordinate singularity at $x=1$
becomes an event horizon, $r=2m$, and a true spacetime
singularity at $r=0$ is hidden by the event horizon.
When $\Delta^2\neq1$, two poles, namely $y=1$ and $y=-1$, 
on the singular surface at $x=1$ have directional properties
as shown below.

In the scalar-tensor-Weyl solution,
a scalar polynomial of the curvature tensor
can be calculated as
\begin{eqnarray}
R^{\mu\nu\sigma\rho} R_{\mu\nu\sigma\rho} & = & 
\frac{4}{\sigma^4}
\left(x-1 \right)^{2\left(\delta-\Delta^2-1 \right)}\left(x+1 \right)^{-2\left(\delta+\Delta^2+1 \right)} 
\left(x^2-y^2 \right)^{2 \Delta^2-3} \times
   \nonumber   \\ 
& &
\left[
-4 \Delta^4\delta^2 \left(1-y^2 \right) 
+4\Delta^4\delta x \left(1-y^2 \right) 
-3 \Delta^4 \left(x^2-y^2 \right)  \right.   \nonumber   \\
& &
+8\Delta^2\delta^3 x \left(1-y^2 \right)    
-2\Delta^2\delta^2  \left(-4+3y^2-6y^2x^2+7x^2 \right) \nonumber   \\
& &
-4\Delta^2\delta x \left( 1+y^2-2x^2 \right) 
-7\delta^4 \left(x^2-y^2 \right)  \nonumber   \\
& &
\left.
-8 \delta^3 x \left(1+y^2-2x^2 \right)
-4\delta^2 \left(1-y^2-3x^2+3x^4\right)
\right].
\label{pol} 
\end{eqnarray}
As a simple example, we consider a path on the
$(x,y)-$plane given by $y=x-\left(x-1 \right)^\alpha$, where $\alpha$ is 
a constant such that $0<\alpha<1$.
This condition on the index, $\alpha$, ensures that 
inequalities, $x>1$ and $y<1$, are satisfied  
along the path.
Then one finds that
the polynomial (\ref{pol}) along the path  
is expressed 
as the product of some regular function of $(x,y)$ 
and the following factor,
\begin{eqnarray}
\eqalign{
&\left(x-1 \right)^{2\left(\delta-\Delta^2-1 \right)}
\left(x-y \right)^{2 \Delta^
2-3}=\left(x-1 \right)^K,\cr
&K\equiv
\left(2\Delta^2-3\right)\alpha -2 \left(\Delta^2-\delta+1 
\right). \cr}
\label{polpart} 
\end{eqnarray}
While one finds that the polynomial (\ref{pol}) 
diverges along the path when $K<0$,
one does not meet any spacetime singularity
along the path when $K\geq0$.
One notes that 
the inequality, $\Delta^2-\delta+1>0$, always holds
because of the definition of $\Delta$:~
$\Delta\equiv\sqrt{\delta^2+d^2}\geq|\delta|$.
The sign of the index, $K$, therefore depends on the sign of 
the coefficient, $2\Delta^2-3$, as follows:

\

\begin{description}
\item[(a)] The case of $2\Delta^2-3\le 0$\\
The index, $K$, always becomes negative for any positive value
of $\alpha$, accordingly
the pole, $(x, y)=(1, 1)$, is a spacetime singularity.

\item[(b)] The case of $2\Delta^2-3>0$\\
When $\alpha \ge\alpha_0\equiv 
1-\left(2\delta-5\right)/\left(2\Delta^2-3\right)>0$,
$K\geq0$ holds, and one does not meet a spacetime singularity
at the pole, $(x, y)=(1, 1)$, along the path.
\end{description}

\

We now summarize the conditions on the parameters, $\Delta$ and $\delta$,
such that the index, $K$, can be positive.
When the inequality, $\delta<2/5$, holds,
the parameter, $\alpha_0$, defined for the above case (b) 
always becomes greater than unity. 
Since $0<\alpha<1$, 
two conditions, $\Delta^2>3/2$ and $\delta\geq5/2$, 
must hold in order for $K$ to be positive.
 From the another point of view,
one can think of $\alpha_0$ as a function of $\Delta$ and $\delta$. 
Under the above conditions, $\Delta^2>3/2$ and $\delta\geq5/2$,
$\alpha_0$ is a monotonic function 
of $\Delta$ and $\delta$ and has 
a minimal value as $\alpha_0=\left(1+\sqrt{19}\right)/6$ 
for $\Delta=\delta=\left(5+\sqrt{19}\right)/2$. 
Consequently, 
the index, $K$, is always negative
for any values of $\Delta$ and $\delta$ 
in the case of 
$0<\alpha < \left(1+\sqrt{19}\right)/6$.

\

\section{Analytic results on null geodesics 
and optical scalars}

\

We study the properties of
irrotational null geodesics and optical scalars
in the scalar-tensor-Weyl solution by analytic approaches.
Hereafter, we consider the parameter, $\delta$, to be
non-negative such that $m\equiv\delta\sigma\geq0$ holds.
Accordingly, we have $0\leq\delta\leq\Delta$ in
scalar-tensor theories of gravity and $0\leq\delta=\Delta$ in
general relativity.

\

\subsection{Null geodesics} 


\

Because of the conformally invariant nature of null geodesics,
we consider null geodesics on the spacetime
with the metric, $g_{\mu\nu}\equiv A^{-2}\hat{g}_{\mu\nu}$. 
Hereafter, a hat denotes
geometrical quantities associated with $\hat{g}_{\mu\nu}$,
and the corresponding geometrical quantities associated 
with $g_{\mu\nu}$
are denoted without a hat.
Let $\hat{k}^\mu=dx^\mu/dv$ be
a null geodesic tangent associated with $\hat{g}_{\mu\nu}$,
where $v$ is the affine parameter.
Then a null geodesic tangent, $k^\mu=dx^\mu/d\lambda$, 
associated with $g_{\mu\nu}$ is given by
\begin{equation}
d\lambda=A^{-2}dv,~~k^\mu=\frac{dx^\mu}{d\lambda}=A^2\hat{k}^\mu.
\label{ge2}
\end{equation}
The null vectors,  $\hat{k}^\mu$ and $k^\mu$, satisfy the following
geodesic equations,
\begin{equation}
\hat{k}^\alpha\hat{\nabla}_\alpha\hat{k}^\mu
=
k^\alpha\nabla_\alpha k^\mu=0.\label{geadd}
\end{equation}

Since it is difficult to study generic
null geodesics analytically,
we consider specific null geodesics 
which are always on the equatorial plane, $y=0$.
In this case, the geodesic equations are reduced to 
\begin{equation}
\begin{array}{l}
\dot{W}^2=1-\dfrac{h^2}{\sigma^2 \left(x^2-1\right)}
\left(\dfrac{x-1}{x+1}\right)^{2\delta}
\equiv 1-V(x), \\
\dot{t}=\left(\dfrac{x+1}{x-1}\right)^{\delta}, \ \ \ 
\dot{\phi}={\displaystyle\frac{h}{\sigma^2 (x^2-1)}
\left(\dfrac{x-1}{x+1}\right)^{\delta}},
\end{array}
\label{null2}
\end{equation}
where a dot denotes a differentiation 
with respect to the affine parameter, 
$\lambda$, a constant, $h$, is an impact parameter
and $W$ is a function of $x$ defined by
\begin{equation}
\dfrac{dx}{dW}=\dfrac{1}{\sigma}
\left(\dfrac{x^2}{x^2-1}\right)^{\frac{\Delta^2-1}{2}}.
\label{null3}
\end{equation}
One of the particular orbits of interest is
a circular orbit\cite{7}, and it is determined by the conditions 
that $\dot{W}=0$ and $dV/dx=0$. 
These conditions are reduced to $x=2\delta$
when $\delta>1/2$, and the circular orbit
cannot exist when $\delta\leq1/2$.
As will be shown later, $\delta=1/2$ is a critical value
such that many geometrical quantities for $\delta<1/2$
show very different properties compared with those for $\delta>1/2$.

In the asymptotic region such that $r>>m$ (see Section 3.2),
the geodesic equation can be approximated by 
\begin{equation}
\left( \frac{dr}{d \lambda} \right)^2 =
1-\frac{h^2}{r^2}\left(1-\frac{2m}{r}\right)
+\left(1-\frac{h^2}{r^2}\right)
\left(\frac{\Delta^2}{\delta^2}-1\right)\frac{m^2}{r^2}
+O\left(\frac{m^3}{r^3}\right),
\end{equation}
which is compared with the following exact result
in the Schwarzschild spacetime,
\begin{equation}
\left( \frac{dr}{d \lambda} \right)^2 =
1-\frac{h^2}{r^2}\left(1-\frac{2m}{r}\right).
\end{equation}

With the reduced null geodesics equation (\ref{null2}), 
a deflection angle, $\alpha$, of the light path is obtained as
\begin{eqnarray}
\alpha \left(x_o \right) &=& 2\int_{x_o}^{\infty}
\frac{\displaystyle
\left(\frac{x^2}{x^2-1}\right)^{\frac{1-\Delta^2}{2}}}
{\displaystyle\sqrt{x^2-1}
\sqrt{\frac{x^2-1}{x_o^2-1}\left(\frac{x+1}{x-1}\right)^{2\delta}
\left(\frac{x_o+1}{x_o-1}\right)^{-2\delta} -1  
}}\,dx-\pi,
\label{def}
\end{eqnarray}
where a parameter, $x_o$, is related to the impact parameter, $h$, by
\begin{eqnarray}
h&=&\sigma \sqrt{x_o^2-1} \left( \frac{x_o+1}{x_o-1} \right)^\delta.
\label{h}
\end{eqnarray}
Since a deflection angle can be defined only for the scattering orbit,
the allowed range of $x_o$ is given by
$x_o > 2\delta$ for $\delta > 1/2$
and $x_o > 1$ for $\delta \le 1/2$, accordingly
the possible range of the impact parameter is given by
\begin{eqnarray}
h > h_*=\left\{
\begin{array}{ll}
{\displaystyle
2m\sqrt{1-\frac{1}{4\delta^2}} 
\left( \frac{2\delta+1}{2\delta-1} \right)^\delta}  
&  \left( \delta > 1/2\right) \\
4m &  \left( \delta = 1/2 \right)\\
0 &  \left( \delta < 1/2 \right),
\end{array}\right.~~m=\delta\sigma\geq0. 
\label{ho}
\end{eqnarray}
Specific values of $h_*/2m$ are, for example,
2.161, 2.543, 2.598, 2.667 and 2.690 for
$\delta =$0.53, 0.7, 1.0, 1.5 and 2.0, respectively.  
The coordinate, $x$, decreases to 
the ``perihelion'', $x=x_o$, and
then increases along the scattering orbit. 

We obtain an analytic form of the deflection angle,
$\alpha(x_o)$, in the following specific cases.

\

\begin{description}

\item[(a)]~$\delta=0,~\Delta^2=2N~~(N=1,2,3,\cdot\cdot\cdot)$
\begin{equation}
\begin{array}{rcl}
\alpha(x_o)&=&{\displaystyle\pi\left\{
\sqrt{1-\frac{1}{{x_o}^2}}
\sum_{p=0}^{N-1}\frac{(-1)^p(2p-1)!!(N-1)!}
{2^p(N-p-1)!(p!)^2{x_o}^{2p}}
              -1\right\}}\\
&=&{\displaystyle\pi\left\{
\sqrt{1-\frac{1}{{x_o}^2}}\left(
1-\frac{N-1}{2{x_o}^2}+\frac{3(N-1)(N-2)}{16{x_o}^4}
-\cdot\cdot\cdot
\right)
-1\right\}}\\
&\sim&
\begin{cases}
-\pi&(x_o\sim1)\\
{\displaystyle -\frac{\pi\Delta^2}{4{x_o}^2}}&(x_o>>1).
\end{cases}
\label{alpha1}
\end{array}
\end{equation}
One notes that it is the case of $m=0$.
The deflection angle is a negative,
monotonously increasing function of $x_o$
for any value of $N$, as is depicted in Figure 2a.
In addition, the deflection angle can be regarded as a monotonously
decreasing function of $N$.

\item[(b)]~$\delta=1/2,~\Delta^2=4L+2~~(L=0,1,2,\cdot\cdot\cdot)$

\begin{description}

\item[(i)]~$L=0$

\begin{equation}
\begin{array}{rcl}
\alpha(x_o)&=&{\displaystyle
\pi\left\{
\frac{x_o+1}{\sqrt{x_o(x_o+2)}}\left[
\frac{2}{\pi}\sin^{-1}\frac{1}{x_o+1}
+1\right]-1
\right\}}\\
&\sim&
{
\begin{cases}
{\displaystyle \pi\left(\frac{8\sqrt{3}}{9}-1\right)>0}&(x_o\sim1)\\
{\displaystyle \frac{2}{x_o}
-\left(2-\frac{\pi}{2}\right)\frac{1}{{x_o}^2}}&(x_o>>1)~.
\end{cases}
}
\label{alpha2b}
\end{array}
\end{equation}

\item[(ii)]~$L\geq1$

\begin{equation}
\begin{array}{rcl}
\alpha(x_o)&=&{\displaystyle 2\sqrt{x_o(x_o+2)}
\left[\frac{x_o+1}{x_o(x_o+2)}\right]^{4L+1}I_L(x_o)-\pi,}\\
I_L(x_o)&=&{\displaystyle\sum_{n=0}^{L}\sum_{k=0}^{2n}
\frac{(2L)!}{(2L-2n)!(2n-k)!k!}\eta(x_o)^{2(L-n)}
\xi(x_o)^kA_{\ell}(x_o)}\\
&~&{\displaystyle
+\sum_{n=0}^{L-1}\sum_{k=0}^{2n+1}
\frac{(2L)!}{(2L-2n-1)!(2n-k+1)!k!}\eta(x_o)^{2(L-n)-1}
\xi(x_o)^kB_\ell(x_o),}\\
A_p(x_o)&=&{\displaystyle\frac{(2p-1)!!}{(2p)!!}\left\{
\frac{\pi}{2}+\sin^{-1}\frac{1}{x_o+1}\right.}\\
&~&{\displaystyle~~~~~~~~~~~~~~~~~~\left.
-\frac{\sqrt{x_o(x_o+2)}}{x_o+1}\sum_{r=0}^{p-1}
\frac{(2p-2r-2)!!}{(2p-2r-1)!!}\frac{1}{(x_o+1)^{2p-2r-1}}
\right\},}\\
B_p(x_o)&=&{\displaystyle
\frac{(2p)!!}{(2p+1)!!}\frac{\sqrt{x_o(x_o+2)}}{x_o+1}
\sum_{r=0}^p\frac{(2p-2r-1)!!}{(2p-2r)!!}
\frac{1}{(x_o+1)^{2p-2r}},}
\label{alpha2a}
\end{array}
\end{equation}
where $\eta(x_o)=2/(x_o+1),~\xi(x_o)=1-2x_o-{x_o}^2$~and
$\ell=L+n-k$. 

\end{description}

\item[(c)]~$\delta=1/2,~\Delta^2=4L+4~~(L=0,1,2,\cdot\cdot\cdot)$
\begin{equation}
\begin{array}{rcl}
\alpha(x_o)&=&{\displaystyle-2\sqrt{x_o(x_o+2)}
\left[\frac{x_o+1}{x_o(x_o+2)}\right]^{4L+3}J_L(x_o)-\pi,}\\
J_L(x_o)&=&{\displaystyle\sum_{n=0}^{L}\sum_{k=0}^{2n+1}
\frac{(2L+1)!}{(2L-2n)!(2n-k+1)!k!}\eta(x_o)^{2(L-n)}
\xi(x_o)^kA_{\ell+1}(x_o)}\\
&~&{\displaystyle
+\sum_{n=0}^{L}\sum_{k=0}^{2n}
\frac{(2L+1)!}{(2L-2n+1)!(2n-k)!k!}\eta(x_o)^{2(L-n)+1}
\xi(x_o)^kB_\ell(x_o),}
\label{alpha3}
\end{array}
\end{equation}
where $\ell~=~L+n-k$.
The functions,
$\eta(x_o)$,~$\xi(x_o)$,~$A_p(x_o)$~and~$B_p(x_o)$,~are 
the same as those of the case (b).
\end{description}

\

The above two cases, (b) and (c), are grouped into 
the single case that
$\delta=1/2$ and $\Delta^2=2N$, where $N$ is a positive integer.
We depict the deflection angle for several values of $N$
in Figure 2b. It should be noted that,
unless $N$ is unity,
the deflection angle for small values of the impact parameter, $h$,
can become a negative, increasing function of $h$
in marked contrast with the case of the Schwarzschild spacetime.

The appearance of a negative deflection angle indicates 
``reflection'' of a light path, and we will later study under
which conditions the light reflection occurs.

\
\clearpage
  \begin{figure}[t]
 \begin{tabular}{cc}
  \includegraphics[width=.45\textwidth]{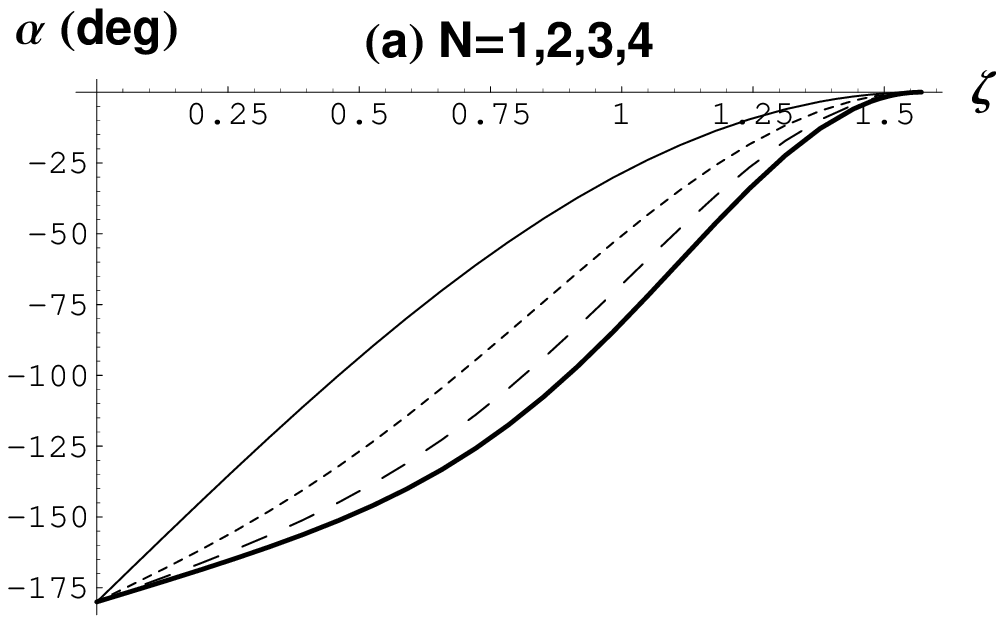} &
  \includegraphics[width=.45\textwidth]{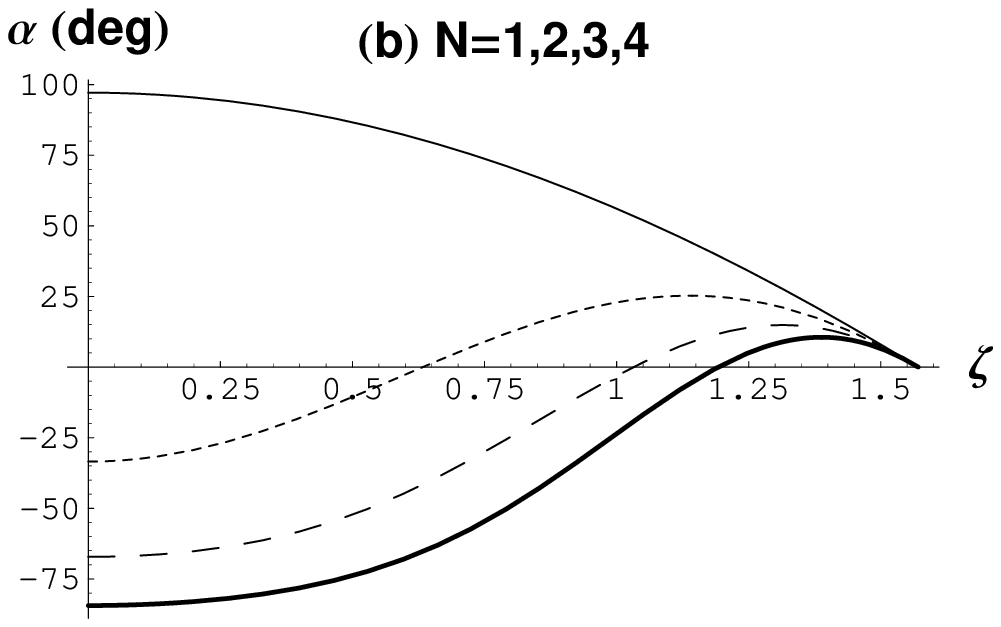} 
      \label{fig.2}
   \end{tabular}
\caption{
The deflection angle, $\alpha(x_o)$,
is shown for some specific values of the parameters, $(\delta,\Delta)$, 
where $\zeta$ is defined by $x_o=1/\cos\zeta$, 
and $\alpha(x_o)$ is given in the unit of degree. 
The model parameters
are (a) $\delta=0$, $\Delta^2=2N$ and
(b) $\delta=1/2$, $\Delta^2=2N$.
We use
a thin line for $N=1$,
a short broken line for $N=2$,
a long broken line for $N=3$ and
a bold line for $N=4$. 
           } 
  \end{figure}
\  

\subsection{Optical scalars} 

\

Since the local geometrical nature of the spacetime is
described by the Riemann curvature, it is important to
investigate the properties of optical scalars of 
the irrotational null geodesic congruence\cite{10}.
Let $\{\hat{\bf E}_{(1)},\hat{\bf E}_{(2)},\hat{\bf E}_{(3)},
\hat{\bf E}_{(4)}\}
=\{\hat{\bf k}, \hat{\bf m}, \hat{\bf t}, \bar{\hat{\bf t}}\}$
be a null tetrad such that
\begin{equation}
\hat{k}^\mu\hat{\nabla}_\mu\hat{E}_{(a)}^\nu=0,~~~
\hat{g}_{\mu\nu}\hat{E}_{(a)}^\mu\hat{E}_{(b)}^\nu=
-\delta^1_{a}\delta^2_{b}-\delta^2_{a}\delta^1_{b}
+\delta^3_{a}\delta^4_{b}
+\delta^4_{a}\delta^3_{b}.
\label{ge3}
\end{equation}
Then the optical scalars, namely, the expansion, $\theta$, and
the complex shear, $\sigma$, are defined by
\begin{equation}
\hat{\theta}=\hat{\nabla}_\mu\hat{k}_\nu\hat{t}^\mu
\bar{\hat{t}}^\nu,~~~
\hat{\sigma}=\hat{\nabla}_\mu\hat{k}_\nu
\bar{\hat{t}}^\mu
\bar{\hat{t}}^\nu.
\label{ge4}
\end{equation}
The optical scalar equations are also conformally invariant,
and we define the following quantities associated 
with $g_{\mu\nu}$,\cite{7}
\begin{equation}
\theta=A^2\hat{\theta}-k^\mu\nabla_\mu\ln A,~~
\sigma=A^2\hat{\sigma},~~t^\mu=A\hat{t}^\mu,~~
\bar{t}^\mu=A\bar{\hat{t}}^\mu.
\label{ge5}
\end{equation}
Then the optical scalar equations become\cite{10,11}
\begin{eqnarray}
& &\frac{d\theta}{d\lambda}+\theta^2+|\sigma|^2
=-\frac{1}{2}R_{\mu\nu}k^\mu k^\nu\equiv-{\cal R},\label{ge6a}\\
& &\frac{d\sigma}{d\lambda}+2\theta\sigma
=-R_{\mu\alpha\nu\beta}k^\mu k^\nu\bar{t}^\alpha \bar{t}^\beta
=-C_{\mu\alpha\nu\beta}k^\mu k^\nu\bar{t}^\alpha \bar{t}^\beta
\equiv F,\label{ge6b}
\end{eqnarray}
where $C_{\mu\alpha\nu\beta}$ is the Weyl curvature.

Dyer has obtained general solutions to 
the optical scalar equations
in the general static, spherically
symmetric spacetime.\cite{11} 
We summarize his results in Appendix C, 
and one finds that the Weyl source-term, $F$, 
and the shear, $\sigma$, can be regarded as real quantities
without loss of generality\cite{11}.
Though the scalar-tensor-Weyl solution is not
spherically symmetric, we can apply Dyer's method
as concerns the null geodesics 
on the equatorial plane, $y=0$\cite{7}.
By defining the new optical scalars, $C_{+}$ and $C_{-}$, as
\begin{equation}
\frac{d}{d\lambda}\ln C_+=\theta+\sigma,~~
\frac{d}{d\lambda}\ln C_-=\theta-\sigma,
\label{dyer1}
\end{equation}
Dyer has obtained the following equations,
\begin{equation}
\frac{d^2C_+}{d\lambda^2}=(-{\cal R}+ F)C_+,~~
\frac{d^2C_-}{d\lambda^2}=(-{\cal R}- F)C_-.
\label{dyer2}
\end{equation}
The equations (\ref{dyer2}) are equivalent to the geodesic deviation
equations, and the variables, $C_{+}$ and $C_{-}$, represent,
respectively,
the size of the major axis and the minor axis of the infinitesimal
image in the Einstein frame. We define the conformal transformation
of $C_\pm$ by $\hat{C}_{\pm}=AC_\pm$ such that this transformation
law is compatible with the others.

The Ricci and Weyl source-terms, ${\cal R}(x)$ and $F(x)$, 
in the Scalar-tensor-Weyl solution on the equatorial plane, $y=0$,
are calculated as (Appendix C, Ref.7)
\begin{eqnarray}
{\cal R}(x) & = &
\dfrac{\Delta^2-\delta^2}{\sigma^2}
\dfrac{1}{x^2(x^2-1)}\left(\dfrac{x^2}{x^2-1}
\right)^{\Delta^2}\left[1-\dfrac{(h/\sigma)^2}{x^2-1}
\left(\dfrac{x-1}{x+1}\right)^{2\delta}\right]\nonumber\\
 & = & \dfrac{\Delta^2-\delta^2}
{(x^2-1)^2}(k^1)^2,~~k^1=\frac{dx}{d\lambda},\label{RFa}\\
F(x)& = &
\dfrac{1}{\sigma^2}\dfrac{1}{x^3(x^2-1)}\left(\dfrac{x^2}{x^2-1}
\right)^{\Delta^2} \times \tilde{F}(x),\nonumber\\
\tilde{F}(x)&=& 
\delta(\Delta^2-1)\nonumber\\
& &+\dfrac{(h/\sigma)^2}
{x^2-1} \left(\dfrac{x-1}{x+1}\right)^{2\delta}
\left\{\delta(\Delta^2-1)-(\Delta^2+2\delta^2)x
+3\delta x^2\right\},\label{RFb}
\end{eqnarray}
where $h$ is the impact parameter as before.
In general relativity,  we have $\Delta=\delta$ and ${\cal R}(x)=0$.
In the asymptotic region such that $x>>1$,
${\cal R}(x)$ and $F(x)$, respectively, approximate to
\begin{eqnarray}
{\cal R}(x) & \sim &
\frac{m^2\left( \Delta^2-\delta^2 \right)}{\delta^2 r^4}
\left[1+\frac{4m}{r}
-\frac{h^2}{r^2}\left(1+\frac{m}{r}\right)
\right],
\label{RR}\\  
F(x)& \sim &
\frac{3mh^2}{r^5}\left(
1-\frac{\Delta^2-\delta^2}{3\delta^2}\frac{m}{r}
+\frac{4\Delta^2+\delta^2-5}{6\delta^2}\frac{m^2}{r^2}
\right)
+\frac{\Delta^2-1}{\delta^2}\frac{m^3}{r^5},
\label{FF}
\end{eqnarray}
in terms of $m/r$, where $x$ is related to $r$ by 
(\ref{x})$\sim$(\ref{chi}) with $y=0$.

It is immediately found that the Ricci source-term, ${\cal R}(x)$, 
satisfies ${\cal R}(x)\ge 0$ and necessarily vanishes 
at the perihelion, $x=x_o$, because $k^1$ vanishes there. 
As for the Weyl source-term, $F(x)$, its value 
is not always positive and can become negative along the light path.
Since the Weyl source-term is mostly dominant 
over the Ricci source-term as will be found by the later numerical study, 
we classify the parameter space, $(\delta,\Delta)$, according to
the qualitative properties of the Weyl source-term as follows.

When $\delta=\Delta=1$, the scalar-tensor-Weyl solution becomes
the Schwarzschild spacetime, and we have
\begin{equation}
F(x)=\frac{3h^2}{\sigma^4(x+1)^5}=\frac{3mh^2}{r^5}.
\label{Fsch}
\end{equation}
One notes that
the Weyl source-term (\ref{Fsch}) has the following properties:

\

\begin{description}

\item[A1]~It is strictly positive definite for $x>x_o$.
 
\item[A2]~It is a monotonously decreasing function of $x$.

\item[A3]~It takes the maximum value at the perihelion, $x=x_0$,
along the scattering orbit.
\end{description}

\

Because of the properties above, one may simply expect
that light rays are affected by the Weyl source-term 
mostly near the perihelion.
For this reason, we first
classify the parameter space, $(\delta,\Delta)$, according to
the value of $F(x_o)$.  
With (\ref{h}) and (\ref{RFb}), one finds that
\begin{eqnarray}
F(x_o)&=&\frac{1}{\sigma^2}\frac{1}{{x_o}^3({x_o}^2-1)}
\left(\frac{{x_o}^2}{{x_o}^2-1}\right)^{\Delta^2}
f(x_o),\nonumber\\
f(x_o)&=&3\delta {x_o}^2-(\Delta^2+2\delta)x_o+2\delta(\Delta^2-1).
\end{eqnarray}
The most intriguing thing is that $F(x_o)$ becomes zero 
for $x_o=x_{F+}$ and $x_o=x_{F-}$, where
\begin{eqnarray}
x_{F\pm} &=& \frac{1}{6\delta}\left\{\left(\Delta^2+2\delta^2 \right)
\pm\sqrt{D} \right\},  \nonumber \\
D &=& \left(\Delta^2-10\delta^2\right)^2
-24\delta^2\left(2\delta-1\right)\left(2\delta+1\right).
\label{D}
\end{eqnarray}
Note that $x_{F\pm}$ and $D$ are, respectively, roots and
a discriminant of the quadratic equation, $f(x_o)=0$.
Since $x_o$ should be real and moreover satisfy the condition, 
$x_o>2\delta$ for $\delta>1/2$ and
$x_o>1$ for $\delta\leq1/2$, we classify
the parameter space, $(\delta,\Delta)$, on the basis
of the number of the conditional roots as follows:

\

\begin{description}
\item[(a)]~The region of $\delta>1/2$\\
The discriminant, $D$, should be positive or zero
in order for $x_{F\pm}$ to be real, and 
this condition is satisfied for the two cases, namely
the case of $\Delta^2\leq{\Delta_-}^2$ 
and the case of $\Delta^2\geq{\Delta_+}^2$, where
\begin{equation}
\Delta_{\pm}^2=10\delta^2 
\pm \sqrt{24\delta^2\left(2\delta-1\right)\left(2\delta+1\right)}.
\label{Dpm}
\end{equation}
For $\delta>1/2$, ${\Delta_+}^2$ and ${\Delta_-}^2$ are
always real and moreover satisfy 
the relation, $0<{\Delta_-}^2<{\Delta_+}^2$.
In this region, a further condition, $x_o>2\delta$, 
should be satisfied for the scattering orbit, and we find that
it holds only for the case, $\Delta^2\geq{\Delta_{+}}^2$,
and is always violated for the case, $\Delta^2\leq{\Delta_{-}}^2$.
We also find that the case of 
$x_{F-}<2\delta<x_{F+}$ does not occur.
The equations, $\Delta={\Delta_+}$ and $\delta=1/2$, define 
the critical lines on the parameter space, $(\delta,\Delta)$, 
on which $x_{F+}=x_{F-}$ and ${\Delta_+}^2 ={\Delta_-}^2$
hold, respectively.
In summary, the region of $\delta>1/2$ is
classified into the following two regions:

\

\begin{description}
\item[(i)]~$\Delta^2<{\Delta_+}^2$\\
We shall refer to the region determined by the conditions,
$\delta>1/2$ and $\Delta^2<{\Delta_+}^2$, 
as the region ${{\rm N}_{\rm A}}$ 
(N: none) because
the equation, $F(x_o)=0$, has no conditional root 
in this region.

\item[(ii)]~$\Delta^2>{\Delta_+}^2$\\
We shall refer to the region determined by the conditions,
$\delta>1/2$ and $\Delta^2>{\Delta_+}^2$, 
as the region II (II: two) because
the equation, $F(x_o)=0$, has two conditional 
roots, $x_{F+}$ and $x_{F-}$, in this region.
\end{description}

\

\item[(b)]~The region of $0<\delta<1/2$ \\
The quadratic equation, $f(x_o)=0$, always has 
two real roots, $x_{F+}$ and $x_{F-}$, because 
its discriminant, $D$, is positive-definite for $0<\delta<1/2$.
In this region, a further condition, $x_o>1$, 
should be satisfied for the scattering orbit, and we find that
$x_{F-}$ is inevitably less than unity.
The condition, $x_{F+}>1$, can be reduced to $\delta<\Delta^2$,
and we have $x_{F+}=1$ on the new critical line
defined by $\delta=\Delta^2$. 
In summary, the region of $0<\delta<1/2$ is
classified into the following two regions:

\

\begin{description}
\item[(i)]~$\delta<\Delta^2$\\
We shall refer to the region determined by the conditions,
$0<\delta<1/2$ and $\delta<\Delta^2$, 
as the region I (I: one) because
the equation, $F(x_o)=0$, has a single conditional root, $x_{F+}$, 
in this region.

\item[(ii)]~$\delta>\Delta^2$\\
We shall refer to the region determined by the conditions,
$0<\delta<1/2$ and $\delta>\Delta^2$, 
as the region ${{\rm N}_{\rm B}}$ because
the equation, $F(x_o)=0$, has no conditional 
root in this region. 
\end{description}

\

\item[(c)]~The critical line, $\delta=1/2$\\
We have a single conditional root, $x_{F+}=2(\Delta^2-1)/3$, for
$\Delta^2>5/2$ and no root for $\Delta^2\leq5/2$.

\item[(d)]~$\delta=0$\\
For any positive value of $\Delta^2$, $F(x_o)$ is negative.
\end{description}

\

The results of the above classification
are shown in Figure 3.

\begin{figure}[ht]
\centerline{\rotatebox{-90}
{\includegraphics[width=.34\textwidth]{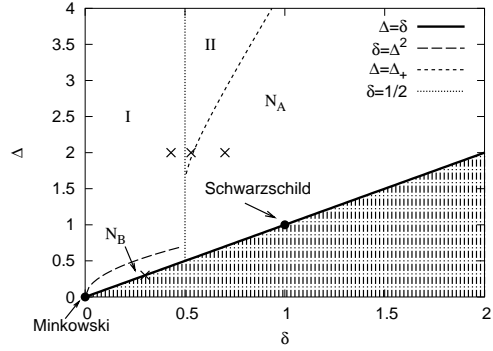}}}
      \caption{
We classify the parameter space, $(\delta,\Delta)$, 
into four regions (${{\rm N}_{\rm A}}$, ${{\rm N}_{\rm B}}$, 
I and  II) 
according to the number of roots of
a quadratic equation, $f(x_o)=0$, with the additional conditions
described in the text.
Horizontal and vertical axes denote 
$\delta$ and $\Delta$, respectively, and 
$\Delta \ge\delta$ in the scalar-tensor-Weyl solution. 
Three critical lines, $\delta=1/2$, $\delta=\Delta^2$
and $\Delta={\Delta_+}$, are shown, where
$\Delta_{+}$ is given by
$\Delta_{+}^2
=10\delta^2 + \delta\sqrt{24\left(2\delta-1\right)
\left(2\delta+1\right)}.$
Two specific points, $(\delta,\Delta)=(0,0)$ 
and $(\delta,\Delta)=(1,1)$, represent 
the Minkowski spacetime and the Schwarzschild 
spacetime, respectively. 
When we show some typical examples
of the numerical results, we will mostly consider
the points denoted by $\times$ 
as standard ones (see Section 5.1).  
}
\label{fig.3}
\end{figure}
  \begin{figure}[ht]
 \begin{tabular}{cc}
  \includegraphics[width=.41\textwidth]{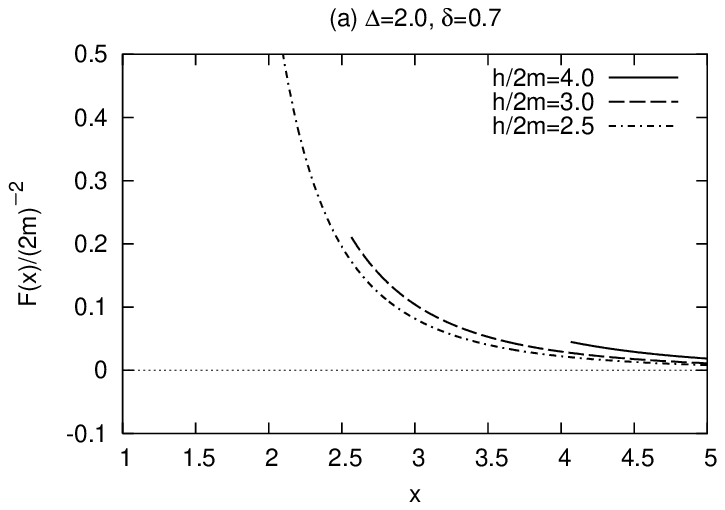} &
  \includegraphics[width=.41\textwidth]{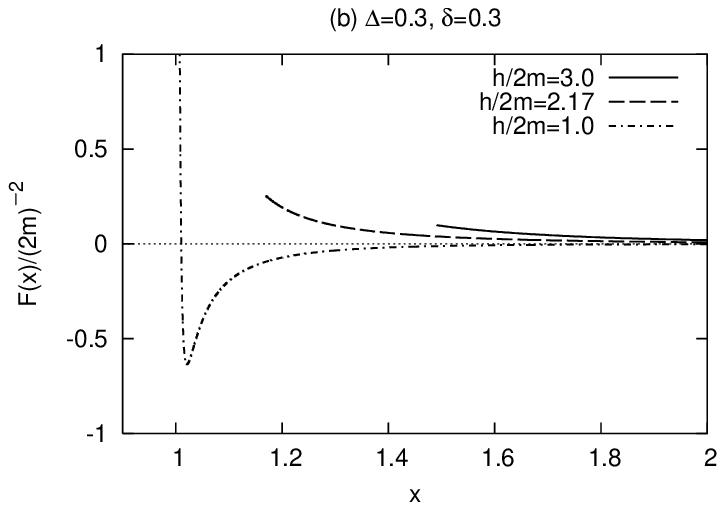} \\
  \includegraphics[width=.41\textwidth]{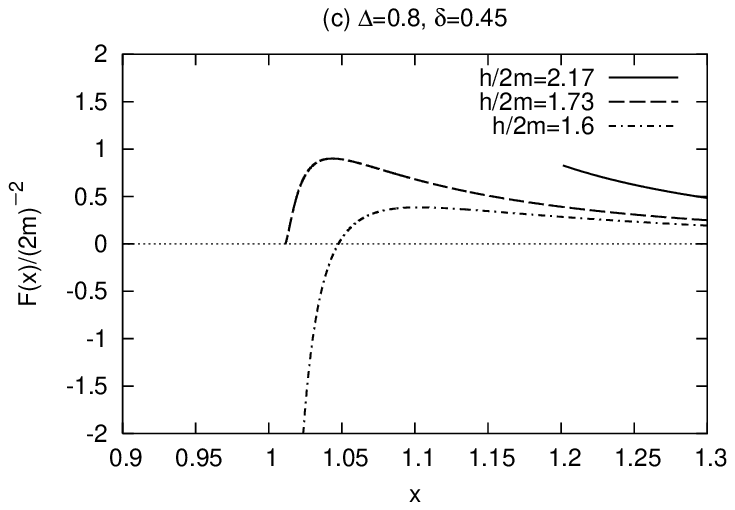} &
  \includegraphics[width=.41\textwidth]{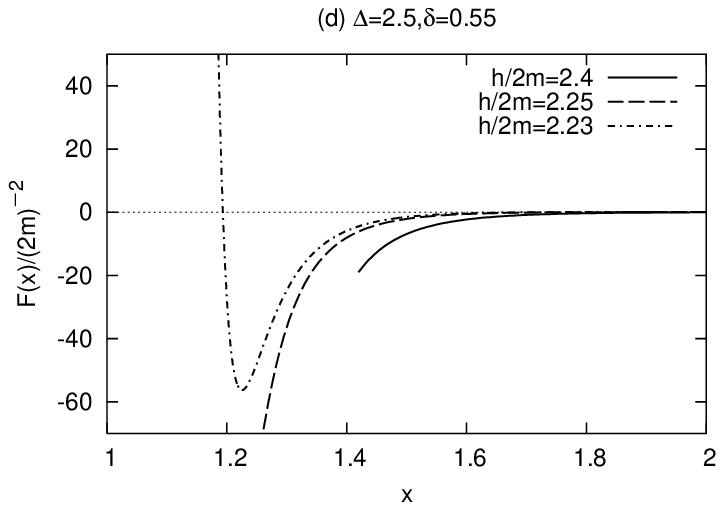}  
      \label{fig.4}
   \end{tabular}
      \caption{
The Weyl source-term, $F(x)$, is shown as a function 
of $x$ ($x_o<x$)
for several values of $\Delta$ and $\delta$:
(a)~$(\delta,\Delta)=(0.7,2.0)$ in the region ${{\rm N}_{\rm A}}$,
(b)~$(\delta,\Delta)=(0.3,0.3)$ in the region ${{\rm N}_{\rm B}}$,
(c)~$(\delta,\Delta)=(0.45,0.8)$ in the region I and
(d)~$(\delta,\Delta)=(0.55,2.5)$ in the region II.
}
  \end{figure}
\clearpage
In Figure 4, we show the Weyl source-term, $F(x)$, as a function of 
$x$ ($x_o<x$) for several values of the parameters, 
$\delta$ and $\Delta$, and the impact parameter, $h$.
While $F(x)$ has similar properties 
as those in the case of the Schwarzschild spacetime 
for large $h$ (see A1$\sim$A3), 
the following quite different properties B1$\sim$B3 can
appear for small $h$. 

\

\begin{description}

\item[B1]~It can be negative, and the maximum number of roots of
the equation, $F(x)=0$, is two for $x>x_o$.

\item[B2]~It can have extremes, and the maximum number 
of the extremes is two for $x>x_o$.
 
\item[B3]~It does not necessarily
take the maximum value at the perihelion, $x=x_o$,
along the scattering orbit.
\end{description}

\

We numerically investigate the function, $F(x)$,
and classify the parameter space, $(\delta,\Delta)$,
by the maximum number of the roots (B1) and
the maximum number of the extremes (B2). 
The results are summarized in Figure 5.
In Figure 5a, the maximum number of the roots is shown
by symbols, $*$ and $+$.
In the region denoted by $*$, there exists at least one
impact parameter, $h$, such that $F(x)=0$ has two roots.
In the region denoted by $+$, $F(x)=0$ cannot have two roots
for any $h$, and there exists at least one
impact parameter, $h$, such that $F(x)=0$ has a single root.   
In the blank region, $F(x)=0$ cannot have a root for any $h$. 
The maximum number of the extremes is shown 
in Figure 5b in the similar
manner as in the case of Figure 5a, namely 
$*$ for two extremes, $+$ for one extreme at the maximum and
blank for no extreme.
One immediately finds close resemblance 
between these classified regions in Figures 5a,b
and the previous one in Figure 3.
We will numerically show
that the classification presented in Figure 3 really
has a close relationship with several properties of light propagation and
gravitational lensing. 
  \begin{figure}[b]
 \begin{tabular}{cc}
  \rotatebox{-90}{\resizebox{45mm}{!}{\includegraphics{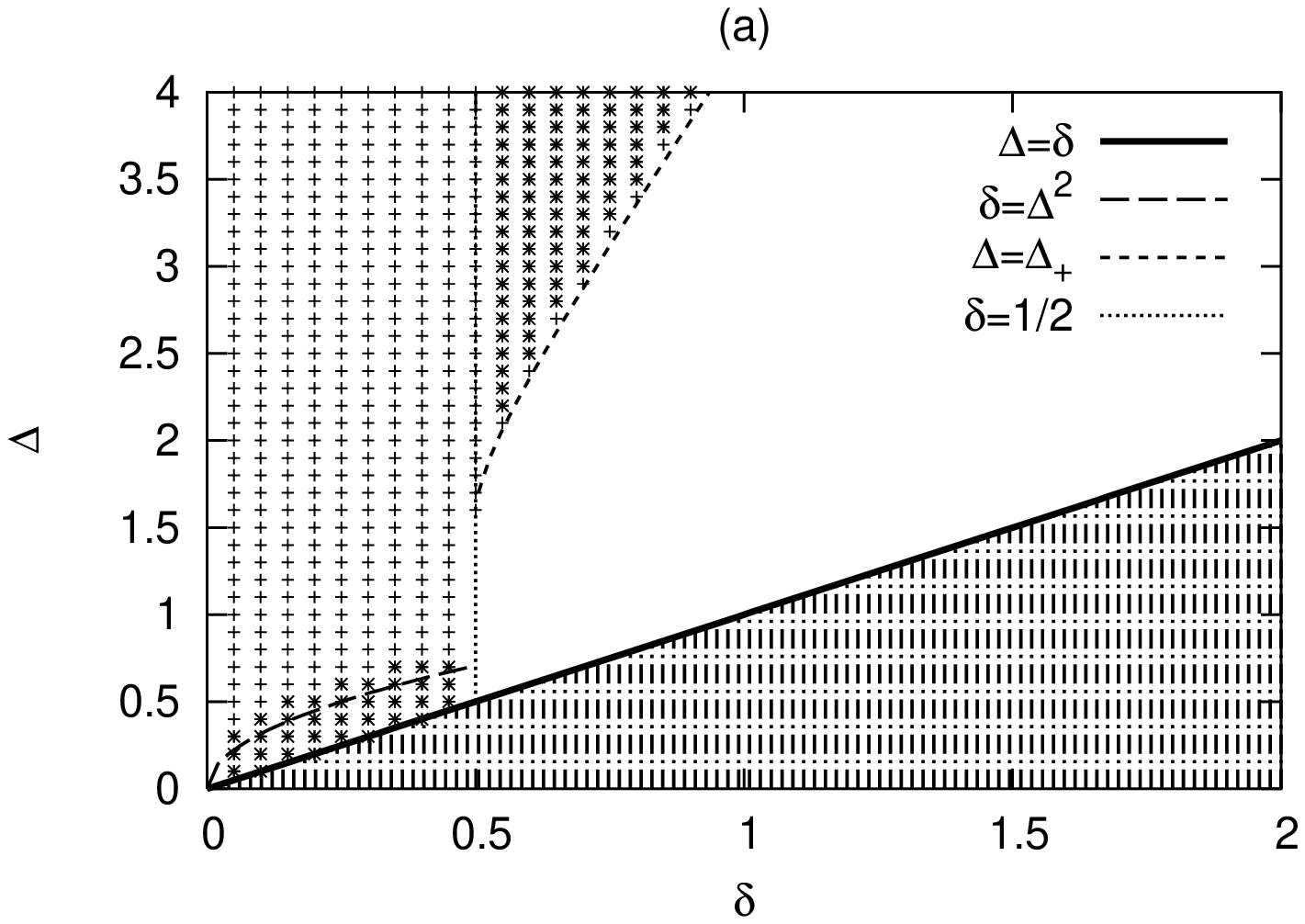}}} &
  \rotatebox{-90}{\resizebox{45mm}{!}{\includegraphics{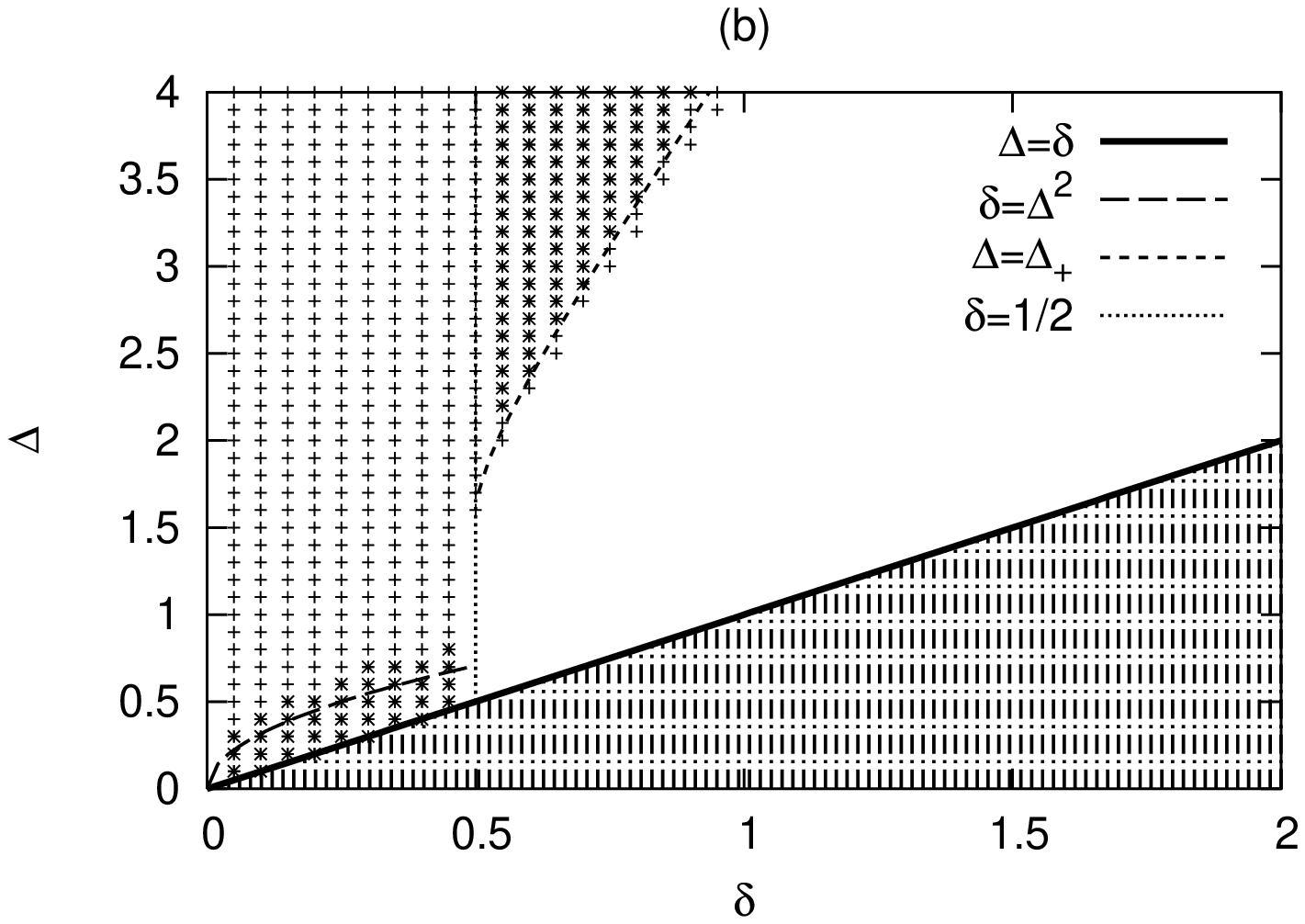}}} 
      \label{fig.5}
   \end{tabular}
      \caption{
We classify the parameter space, $(\delta,\Delta)$, 
according to the qualitative properties of
the Weyl source-term, $F(x)$:
(a) classification by the maximum number of roots of $F(x)=0$
and (b) classification by the maximum number of extremes of $F(x)$.
Further details of these figures are explained in the text.
}
  \end{figure}

\

\section{Numerical results on null geodesics and optical scalars}

\

Now we will show several numerical results 
on null geodesics and optical scalars. 

\

\subsection{Preliminaries}

\

There are three model parameters, $\sigma$, $\delta$ and $\Delta$,
in the scalar-tensor-Weyl solution, and we reduce the number
of the parameters to two by fixing the values of $m\equiv\delta\sigma$.
In other words, $m$ plays the role of a unit of length.
When we show some typical examples
of the numerical results, 
the following values of the model parameters, $(\delta,\Delta)$,
will mostly be considered as standard ones (see Figure 3),
\begin{equation}
(\delta,\Delta)=
\begin{cases}
[{\rm n}_{\rm A}]~~(0.7,2.0)&{\rm in~the~region}~{{\rm N}_{\rm A}}\\
[{\rm n}_{\rm B}]~~(0.3,0.3)&{\rm in~the~region}~{{\rm N}_{\rm B}}\\
[~\rm{i}~]~~(0.43,2.0)&{\rm in~the~region}~{\rm I}\\
[~{\rm ii}~]~~(0.53,2.0)&{\rm in~the~region}~{\rm II}~~.
\end{cases}
\end{equation}

In Section 4.2 we have investigated
the conditional roots of the equation, $F(x_o)=0$, and find 
the single root, $x_{F+}$, in the regions I 
and the two roots, $x_{F+}$ and $x_{F-}$ in the region II. 
These roots, $x_{F+}$ and $x_{F-}$, are related to
the impact parameters, $h_{F+}$ and $h_{F-}$, by
\begin{eqnarray}
h_{F+}&=&\sigma \sqrt{{x_{F+}}^2-1}
         \left( \frac{{x_{F+}}+1}{{x_{F+}}-1} \right)^\delta,\nonumber\\
h_{F-}&=&\sigma \sqrt{{x_{F-}}^2-1}
         \left( \frac{{x_{F-}}+1}{{x_{F-}}-1} \right)^\delta,
\label{hFpm}
\end{eqnarray}
respectively. 
We will repeatedly emphasize the importance of these specific values 
of the impact parameter, $h_{F+}$ and $h_{F-}$.

On the equatorial plane, $y=0$, a spatial part of 
the metric (\ref{st1}) becomes
\begin{equation}
ds^2  =  
\frac{m^2}{\delta^2}\left(\frac{x-1}{x+1}\right)^{-\delta} 
\left[\left(\frac{x^2-1}{x^2}\right)^{\Delta^2-1}
dx^2
+(x^2-1)d\phi^2\right],
\end{equation}
and the equatorial circumference, $L_x$, is given by
\begin{equation}
L_x=2\pi R_x(x),~~
R_x(x)\equiv\frac{m}{\delta}
\left(\frac{x-1}{x+1}\right)^{-\frac{\delta}{2}}\sqrt{x^2-1}.
\end{equation}
On the basis of this geometrical consideration,
we define the dimensionless, locally Cartesian-like 
coordinates, $(X,Y)$, by 
\begin{equation}
X=\frac{R_x(x)}{2m}\cos\phi,~~
Y=\frac{R_x(x)}{2m}\sin\phi, 
\end{equation}
for the sake of later convenience. 

The optical scalar equations should be solved along the corresponding
light path, and therefore, we 
numerically integrate the geodesic equations and
the optical scalar equations simultaneously under 
the initial conditions as follows:

\

\begin{description}
\item[(a)]~~The direction of the $X$-axis is chosen such that
a light ray initially propagates in the opposite
direction to the $X$-axis, 
namely from ``right'' to ``left'' in  Figure 6.

\item[(b)]~~The affine parameter, $\lambda$, is chosen
such that $\lambda=0$ at the initial time.

\item[(c)]~~Let $x_i$ be an initial value 
of the coordinate, $x$, of the null geodesic at $\lambda=0$.
We should determine $x_i$ such that 
an initial circumference radius, 
$R_i\equiv R_x(x_i)$, is an approximately 
parameter-independent value.  For $x_i=100\delta$, we have
\begin{equation}
\eqalign{
R_i/m&=\frac{1}{\delta}
\left(\frac{x_i-1}{x_i+1}\right)^{-\frac{\delta}{2}}
\sqrt{{x_i}^2-1}\cr
&=\frac{x_i}{\delta}\left\{
1+\frac{\delta}{x_i}+{\cal O}\left(
\frac{\delta^2}{{x_i}^2}\right)
\right\}
\sim101+{\cal O}\left(10^{-2}\right).\cr}
\end{equation} 
Unless otherwise noted, we let a value of $x_i$
be $100\delta$. 

Along the scattering orbit, $x$ decreases from $x_i$ 
to its minimum values, $x_o$, 
and then increases to $x_i$. When $x$ becomes $x_i$ again 
at $\lambda=\lambda_f$, numerical integrations are terminated.
We shall call $\lambda=\lambda_f$ the final time.
Under this geometrical configuration,
we define the Einstein radius, $h_E$, by
\begin{equation}
\frac{h_E}{2m}\equiv\sqrt{\frac{R_i}{2m}}
\sim7.
\end{equation}
The Einstein radius, $h_E$, is usually
defined in the weak gravity region such that $m/h_E<<1$, 
and the above estimate of $h_E$ is
compatible with the assumption 
that the gravity is comparatively weak even at the Einstein radius.

\item[(d)]~~As for the optical scalers, their
initial conditions at $\lambda=0$ are chosen as
\begin{equation}
C_+=C_-=0,~~\frac{dC_+}{d\lambda}=\frac{dC_-}{d\lambda}=1,
\end{equation} 
such that we have $C_+=C_-=\lambda$ in the Minkowski spacetime. 
\end{description}

\

Now we should remind ourselves that the optical scalars
are conformally variant as $\hat{C}_\pm=AC_\pm$.
We shall therefore define a conformally 
invariant optical scalar, $C$, by
\begin{equation}
C\equiv\frac{C_+-C_-}{C_+}
=\frac{\hat{C}_+-\hat{C}_-}{\hat{C}_+}.
\end{equation} 
Since $\hat{C}_{+}$ and $\hat{C}_{-}$ represent
the size of the major axis and the minor 
axis of the infinitesimal image, respectively, 
the new variable, $C$, is a measure of the image distortion rate.
When either $C_+$ or $C_-$ vanishes at the final time 
for some value of the impact parameter, $h$, 
we shall refer to this specific value of the impact parameter
as a caustic point.
It is important to note that we have two kinds of
caustic points, namely the caustic point associated
with $C_-$ and the one associated with $C_+$. 
At the caustic point, we have 
$C\rightarrow1$ for $C_-\rightarrow0$ and
$|C|\rightarrow\infty$ for $C_+\rightarrow0$.

\

\subsection{Null geodesics and the deflection angle}

\

We show the trajectories of light rays
on the $(X,Y)$-plane in Figure 6.
The model parameters, $(\delta,\Delta)$, are chosen 
to be the standard ones and the impact parameter, $h$,
is chosen as (a) $h/2m=3.0$ and (b) $h/2m=2.17$, respectively. 
For the case (b), the light path in the case [${\rm n}_{\rm A}$] fails
to be a scattering orbit, and therefore, we omit this case
in Figure 6b.  Figure 6c
is an enlarged picture of the central part of Figure 6b.   
When the impact parameter, $h$, is comparatively large
as in the case of Figure 6a, the gravity acts attractively,
and the effects of the scalar field and
the non-sphericality remain quantitative.
In contrast to that, Figure 6c shows us
that the gravity can act repulsively for smaller values of $h$,
as has already been predicted 
by the existence of the negative deflection angle
in the previous section. 

  \begin{figure}[b]
 \begin{tabular}{ccc}
  {\includegraphics[width=.45\textwidth]{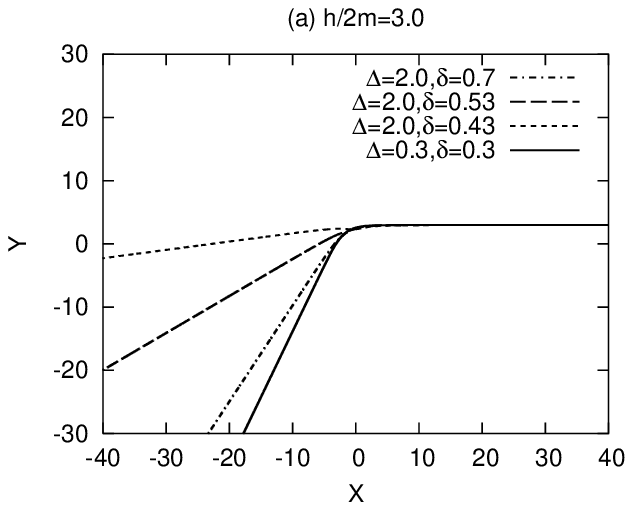}} \\
  {\includegraphics[width=.45\textwidth]{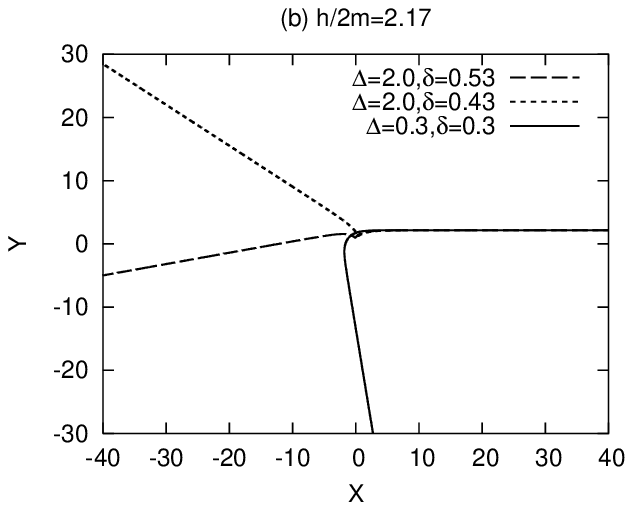}} &
  {\includegraphics[width=.45\textwidth]{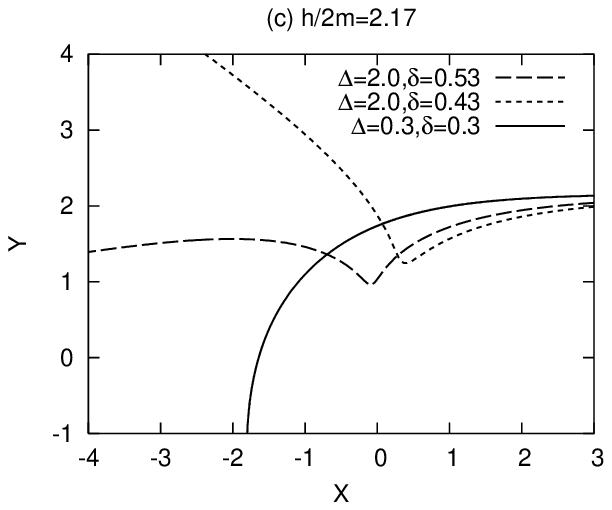}} 
      \label{fig.6}
   \end{tabular}
     \caption{
The trajectories of light rays are shown on the $(X,Y)$-plane
for (a)~$h/2m=3.0$ and (b) $h/2m=2.17$. The enlarged central part
of (b) is shown in (c).
             }
  \end{figure}

The appearance of the repulsive trajectories
explicitly proves that
``reflection'' of a light path can actually occurs, and 
we now investigate the conditions for that.
By the simple geometrical consideration, 
one finds that the situation of $d/d\lambda(dY/dX)>0$
inevitably occurs along a reflecting, 
counterclockwise light path around the singularity at $x=1$.
We succeed in reducing this condition to the occurrence of
$p(x)<0$, where an auxiliary function, $p(x)$, is defined by
\begin{eqnarray}
p(x)&=&\frac{h}{\sigma^4x^3(x^2-1)^2}
\left(\frac{x^2}{x^2-1}\right)^{\Delta^2}\times q(x),\nonumber\\
q(x)&=&\left\{(\delta^2-\Delta^2)x+(\Delta^2-1)\delta\right\}
\left\{h^2\left(\frac{x-1}{x+1}\right)^{2\delta}
-\sigma^2(x^2-1)\right\}\nonumber\\
&~&+h^2x\left(\frac{x-1}{x+1}\right)^{2\delta}
\left\{x^2\left(\frac{x^2-1}{x^2}\right)^{\Delta^2}
-x^2+3\delta x-2\delta^2\right\}.
\label{px}
\end{eqnarray}
In particular, we have
$p(x)=3mh^2/r^5>0$ in the Schwarzschild spacetime.

In Figure 7, we numerically compare the previously classified four 
regions, ${{\rm N}_{\rm A}}\sim{\rm II}$, with 
the reflecting region in which
the inequality, $p(x)<0$, can apply.
Though their relationship is halfway close, it is 
an intriguing thing that some parts of
the boundary of the reflecting region coincide with
the critical line, $\delta=1/2$.

  \begin{figure}[t]
 \begin{tabular}{ccc}
 \centerline{\rotatebox{-90}{\resizebox{45mm}{!}{\includegraphics{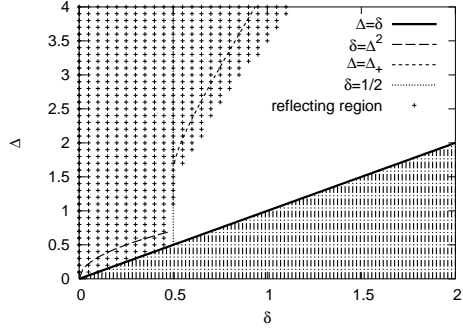}}}}
      \label{fig.7}
   \end{tabular}
      \caption{
The reflecting region is shown by symbols, $+$,
on the classified parameter space, $(\delta,\Delta)$.
}
  \end{figure}

As for the deflection angle, we numerically integrate
(\ref{def}) with (\ref{h}) and depict it as a function
of the impact parameter, $h$, in Figure 8, where 
the marks, $h_1$, $h_2$ and $h_3$, represent
the following specific values of the impact parameters, respectively:
\begin{equation}
(h_1,h_2)=(h_{F-},h_{F+})~~{\rm in~the~case~[~ii~]},
~~~~~~
h_3=h_{F+}~~{\rm in~the~case~[~i~]}.
\end{equation}
We numerically find that the
deflection angle generically has two extremes
near $h=h_{F+}$ and $h=h_{F-}$ in the region II
and a single extreme near $h=h_{F+}$ in the region I. 
Graphs of the deflection angle
in the regions, I and ${{\rm N}_{\rm B}}$, 
have the similar properties as follows:

\

\begin{description}
\item[C1]~~It has a single extreme and the extreme is the maximum.

\item[C2]~~The deflection angle remains to be finite
as the impact parameter approaches to zero.
\end{description}

\

In particular, the properties, C1, are unexpected ones
because of the absence of the characteristic scales, 
$h_{F+}$ and $h_{F-}$, in the region ${{\rm N}_{\rm B}}$.
Let $h_\alpha$ be a new characteristic value
of $h$ at which a graph of the deflection angle
has the maximum.
We numerical find that $h_\alpha$ 
continuously changes beyond the boundaries between
the region I and the region ${{\rm N}_{\rm B}}$
and that $h_\alpha=2m\sim6m$ in the region ${{\rm N}_{\rm B}}$. 
We also find that $h_\alpha$ is a decreasing function of $\delta$ 
and is an increasing function of $\Delta$.

By fixing the values of $\delta$ and $h$, we show the deflection
angle as a function of $\Delta$ in Figure 9
and find that 
the deflection angle is always a decreasing function of $\Delta$. 
We also show the cases in general relativity 
($\Delta=\delta$) for comparison. 

  \begin{figure}[t]
 \begin{tabular}{cc}
  \centerline{{\includegraphics[width=.45\textwidth]
                  {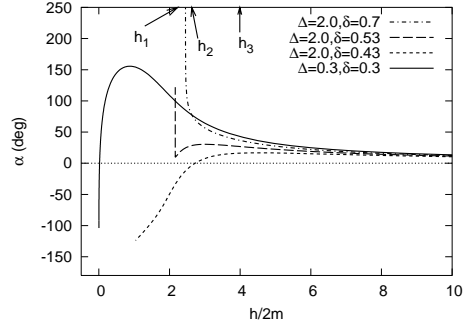}}} 
      \label{fig.8}
   \end{tabular}
     \caption{
The deflection angle, $\alpha$, is shown as a function
of the impact parameter, $h$.
The model parameters are chosen to be the standard ones.
}
  \end{figure}
  \begin{figure}[t]
 \begin{tabular}{cc}
  {\includegraphics[width=.45\textwidth]{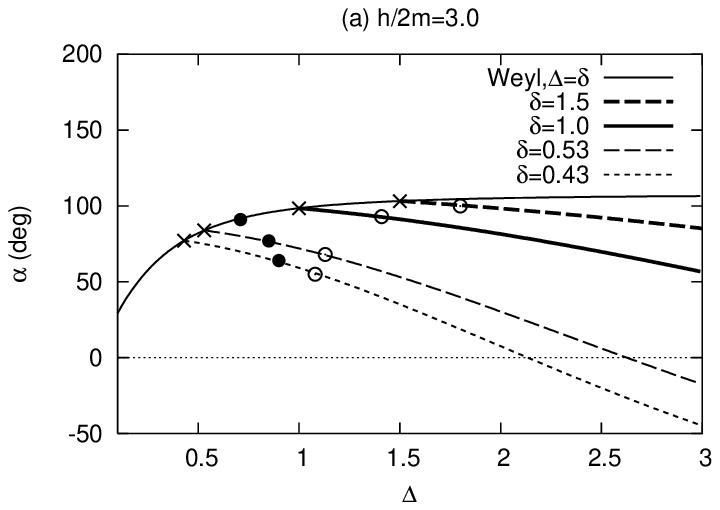}} &
  {\includegraphics[width=.45\textwidth]{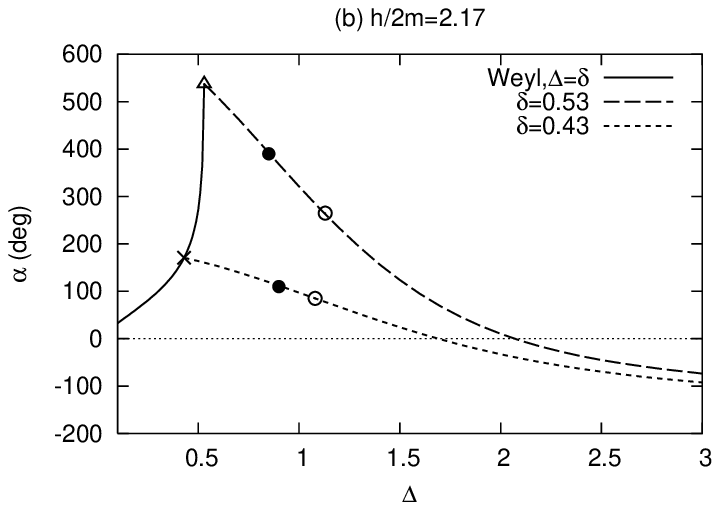}} 
      \label{fig.9}
   \end{tabular}
       \caption{
The deflection angle, $\alpha$, is shown as a function
of the parameter, $\Delta$, for (a) $h/2m=3.0$ and
(b) $h/2m=2.17$.
Thin lines show
the deflection angles in the Weyl solution ($\Delta=\delta$). 
For general cases of $\Delta\geq\delta$, lines corresponding to
the constant $\delta$ have terminal points ($\times$ in the figures)
on the thin line. Open circles and filled circles denote
the data points such that $Q=1$ and $Q=-1$, respectively,
where $Q\equiv(\Delta^2-1)/\delta^2$ is a measure 
of the non-sphericality of the spacetime symmetry. 
An open triangle in (b) denotes the case that the scattering
orbit cannot exist.
}
  \end{figure}

By further numerical investigation on the parameter space,
$(\delta,\Delta)$, we find the following
notable properties of the deflection angle, $\alpha(h)$.

\

\begin{description}
\item[D1]~~While it is positive definite in the Schwarzschild spacetime,
it can be negative in the Scalar-tensor-Weyl solution. 

\item[D2]~~While it is a monotonously decreasing function of $h$ 
in the Schwarzschild spacetime, 
it can have two extremes at the maximum.

\item[D3]~~Since a circular orbit exists when $\delta>1/2$, 
a light ray passing by the circular orbit
can wind around it many times, and the deflection
angle becomes very large.  
In contrast to the cases of $\delta>1/2$, 
it always remains to be finite in the case of $0\leq\delta<1/2$
even when $h$ approaches zero. 
In the case of $\delta=1/2$, we analytically find that
it remains to be finite for $\Delta>1$
and diverges for $\Delta\leq 1$ when $h$ approaches $4m$.
\end{description}

\

We classify the parameter space, $(\delta,\Delta)$, 
according to the number of extremes
of $\alpha(h)$ and compare this new classification
with the previous one, as is shown in Figure 10.
The similar classification according to 
the occurrence of negative deflection angles
is also done, as is shown in Figure 11.
One may immediately note the surprising coincidences
among these results found in Figures 10
and 11.
Though we have not yet succeeded in explaining the reasons
for these coincidences, 
we expect that the classification shown in Figure 3 may have 
some physical significance for gravitational lensing
properties. As for the light reflection and 
the deflection angle, we find that their properties 
in the region ${{\rm N}_{\rm B}}$ is quite similar
to those in the region I.    

Finally, it should be noted that this kind of
the light reflection can be possible even in general relativity.
That is, the light reflection inevitably occurs
even on the line, $\Delta=\delta$, for $\delta<1/2$.
On the other hand, the appearance of the region II must be
an intrinsic effect to the scalar-tensor theories of gravity,
and we expect that the II-like region may always exist
not only in the scalar-tensor-Weyl solution but also
in most of the other solutions. 
While the boundary between the ${{\rm N}_{\rm A}}$-like
region and the I-like region will always be
clear, the boundary between the ${{\rm N}_{\rm A}}$-like
region and the II-like region may, however, be
slightly vague, as has seen in Figures 7, 10 and 11. 

  \begin{figure}[ht]
\begin{minipage}[t]{.49\textwidth}
   \centerline{\rotatebox{-90}{\resizebox{45mm}{!}
            {\includegraphics{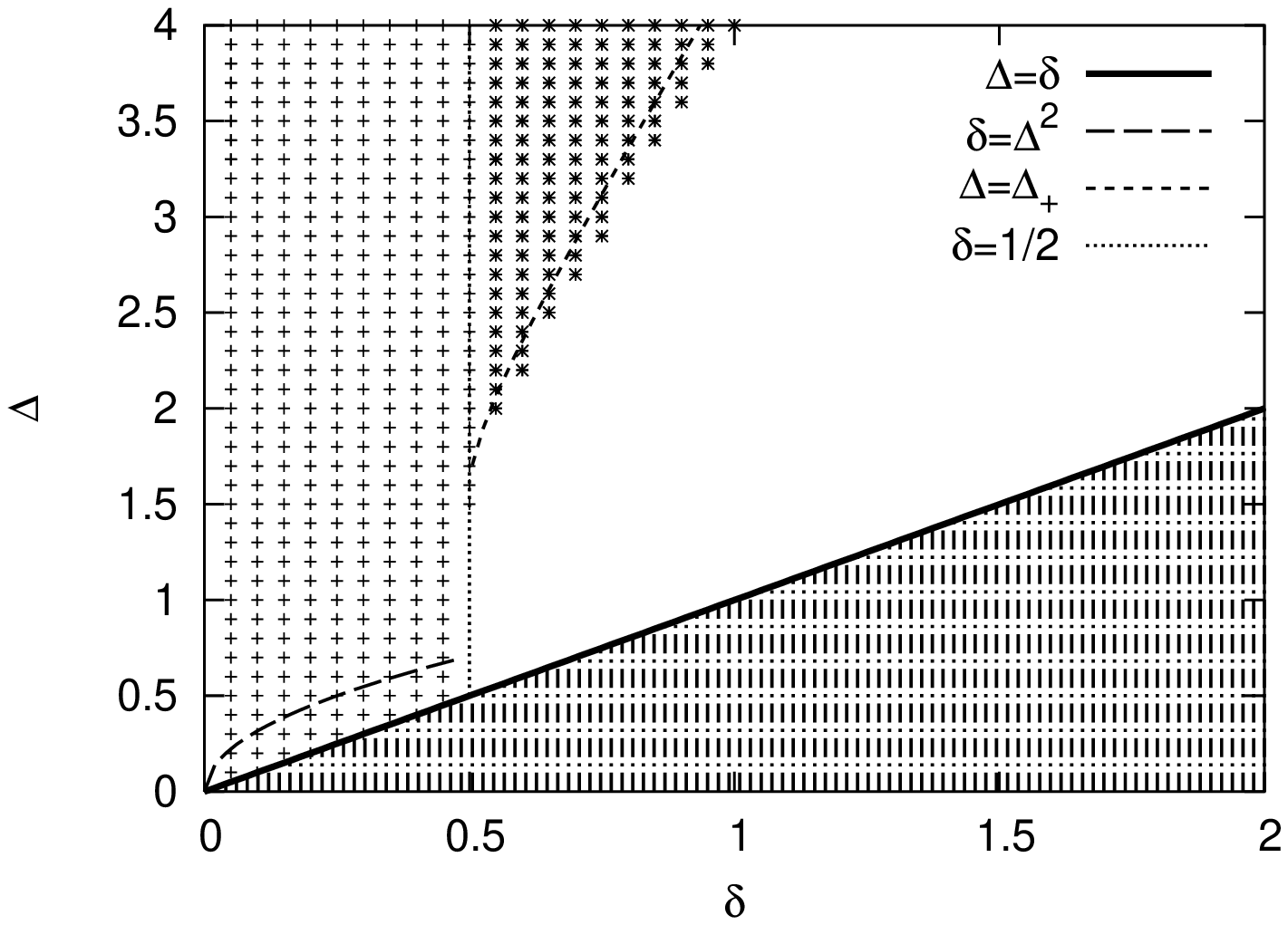}}}} 
      \label{fig.10}
      \caption{
We classify the parameter space, $(\delta,\Delta)$, 
into three regions according to the number of extremes of 
$\alpha(h)$: $*$ for two extremes, $+$ for a single
extreme and $blank$ for no extreme.
}
\end{minipage}
\hfill
\begin{minipage}[t]{.49\textwidth}
  \centerline{\rotatebox{-90}{\resizebox{45mm}{!}
{\includegraphics{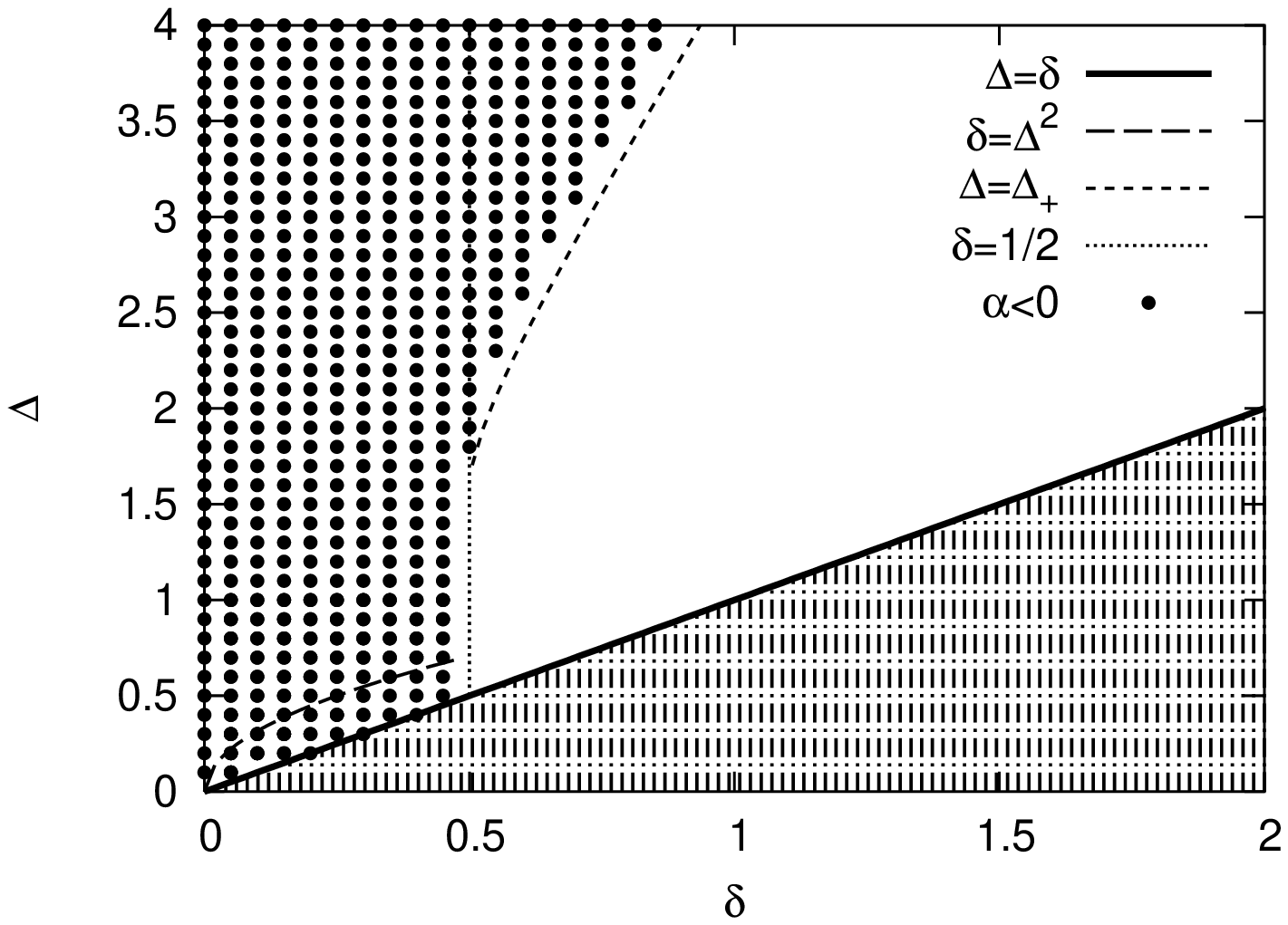}}}} 
      \label{fig.11}
      \caption{
We classify the parameter space, $(\delta,\Delta)$, 
into two regions
according to the occurrence of negative deflection angles.
In the region with filled circles, a deflection angle 
can be negative.
}
\end{minipage}
  \end{figure}

\

\subsection{Optical scalars and the image distortion rate}

\

We numerically solve the optical scalar equations along the light path
by the procedure described in Section 5.1.
In Figure 12, we show (a) $C_+$ and (b) $C_-$ as
functions of the affine parameter, $\lambda$, for several values of
the impact parameters, $h$. 
The model parameters are chosen as
those of the case $\caseNA$, where $(\delta,\Delta)=(0.7,2.0)$.
One must find at first sight that 
gravitational effects
of the source-terms, $F$ and ${\cal R}$, on the light rays
appear as a rapidly changing slope of the graph of 
$C_+$ in the vicinity of $\lambda=50$.
As for $C_-$, the slope changes once for 
$h/2m=$3.0~and~4.0 and twice for $h/2m=2.5$.
This ``bending'' of the graph is found 
in the narrow range of $\lambda$,
and its width, $\Delta\lambda$, is $2\sim3$ for
the case of $C_-$ with $h/2m=2.5$ and
invisibly thin for the other ones.  
In this sense, a picture of the thin lens is applicable, 
and, roughly speaking, the gravitational effects appear
as a bending angle of the graph.    
  \begin{figure}[ht]
 \begin{tabular}{cc}
  {\includegraphics[width=.41\textwidth]{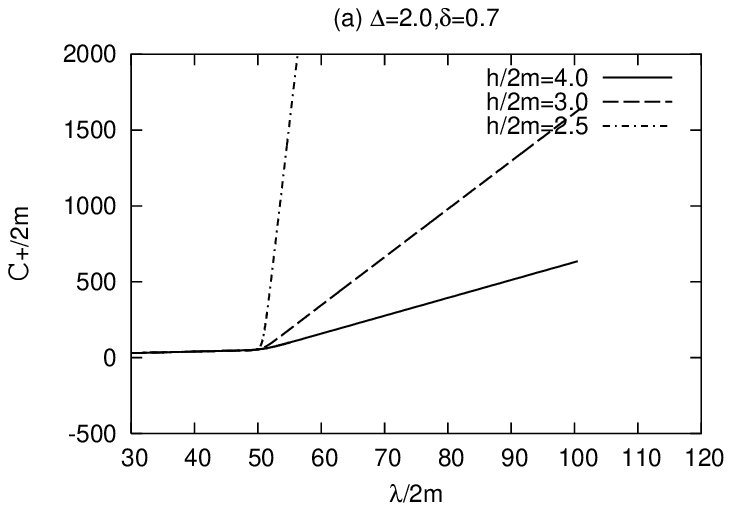}}  &
  {\includegraphics[width=.41\textwidth]{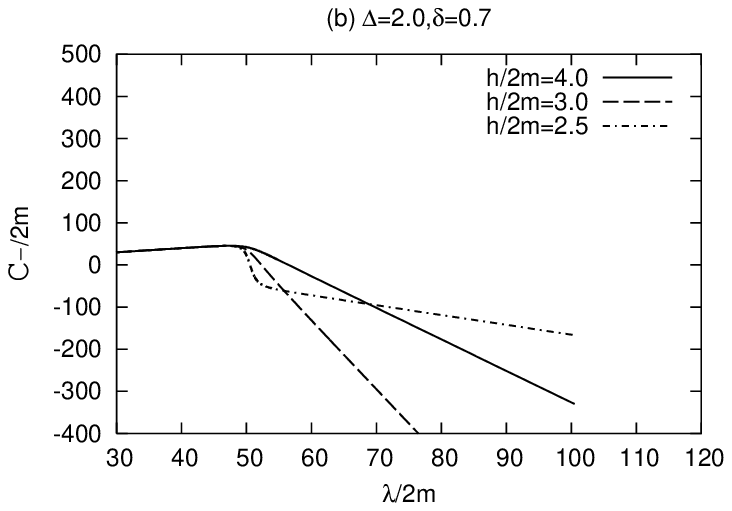}}  
      \label{fig.12}
   \end{tabular}
      \caption{
We show (a) $C_+$ and (b) $C_-$ as
functions of $\lambda$ for several values of
the impact parameters, $h$, as denoted in each figure. 
The model parameters are chosen as
those of the case $\caseNA$, where $(\delta,\Delta)=(0.7,2.0)$.
}
  \end{figure}

In Figures 13, 14 and 15, we show (a) $C_+$ and (c) $C_-$ 
for the standard cases, $\caseNB$, $\caseI$ 
and $\caseII$. 
For comparison, 
the enlarged pictures of these graphs are
also shown in (b) and (d) for $C_+$ and $C_-$, respectively.
In any case, a rough picture of the thin lens is applicable again.
To confirm the forecast that the lenses are thin, 
we show the Ricci and Weyl source-terms as functions 
of the affine parameter, $\lambda$, in Figures 16$\sim$19. 
These thin lenses are, however, divided into 
the following two groups by
their thinness (or thickness).

\

\begin{description}
\item[E1]~A lens is invisibly thin such that the bending angle of
the graph is a moderately varying function of the impact parameter, $h$.

\item[E2]~A lens has a visibly finite thickness, 
$\Delta\lambda$, such that 
one recognizes a comparatively rapid variation of the graph
when light rays goes through the lens.
\end{description}

\

It should be noted that the above classification of lenses
is brought not so much by the intrinsic properties 
of each classified region in the parameter space 
as by the extrinsic factors, especially the impact parameter, $h$,
and the initial distance, $R_i$,
as is explained by the following simple model.

Under the thin lens assumption, we approximate
the optical scalar equations (\ref{dyer2}) by
\begin{equation}
\frac{d^2u_\pm}{d\lambda^2}=
\begin{cases}
0&(|\lambda|>\epsilon)\\
\pm\kappa^2u_{\pm}&(|\lambda|<\epsilon), 
\end{cases}
\label{upm}
\end{equation}
where $\kappa^2$ and $\epsilon\sim\Delta\lambda/2$ are 
an amplitude of the source-term and
a thickness of the lens, respectively.
The affine parameter, $\lambda$, is redefined
such that the center of the thin lens is at $\lambda=0$.
For example,
when the combined source-term, $-{\cal R}+F$, of $C_+$
is negative, the negative sign of (\ref{upm}) is chosen for $C_+$:
\begin{equation}
\frac{d^2{C_+}}{d\lambda^2}=\left(
-{\cal R}+F
\right)C_+~~\rightarrow~~
\frac{d^2u_-}{d\lambda^2}=
\begin{cases}
0&(|\lambda|>\epsilon)\\
-\kappa^2u_{-}&(|\lambda|<\epsilon). 
\end{cases}
\end{equation}
The equation (\ref{upm}) is easily solved
under the initial conditions,
\begin{equation}
u_\pm=0,~~
\frac{du_\pm}{d\lambda}=1
~~{\rm at}~~\lambda=-\lambda_0<0,
\end{equation}
and we find the following properties 
under the condition that
$\epsilon\sim\Delta\lambda/2<<\lambda_0\sim R_i$.

\

\begin{description}
\item[F1]~The case of $\epsilon\kappa<<1$\\
It corresponds to the case of E1.
The bending angle, $\Delta\theta_{\pm}$, of
the graph of $C_\pm$ is well approximated as
\begin{equation}
\Delta\theta_\pm=\pm(\kappa\lambda_0)(2\kappa\epsilon)
\sim\pm\kappa^2\lambda_0\Delta\lambda.
\end{equation}
Note that $|\Delta\theta_\pm|$ 
can be larger than unity for the cases of large values of
$\kappa\lambda_0\sim\kappa R_i$.

\item[F2]~The case of $\epsilon\kappa>1$\\
It corresponds to the case of E2. When light rays
pass through the lens,
$u_-$ oscillates approximately $\epsilon\kappa/\pi$ times,
while $u_+$ exponentially grows as
\begin{equation}
\log u_+\sim\kappa\lambda+
\log\left(\frac{1}{2}\lambda_0{\rm e}^{\epsilon\kappa}\right).
\label{uplus}
\end{equation}
\end{description}

\

  \begin{figure}[ht]
 \begin{tabular}{cc}
  {\includegraphics[width=.41
\textwidth]{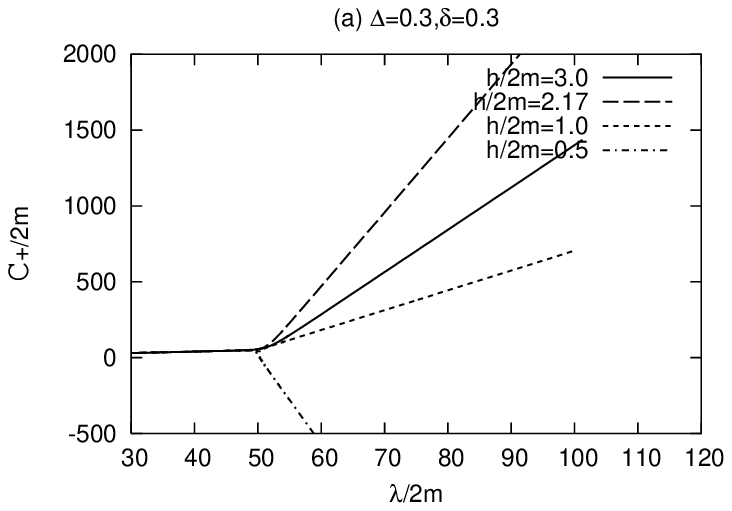}}  &
  {\includegraphics[width=.41
\textwidth]{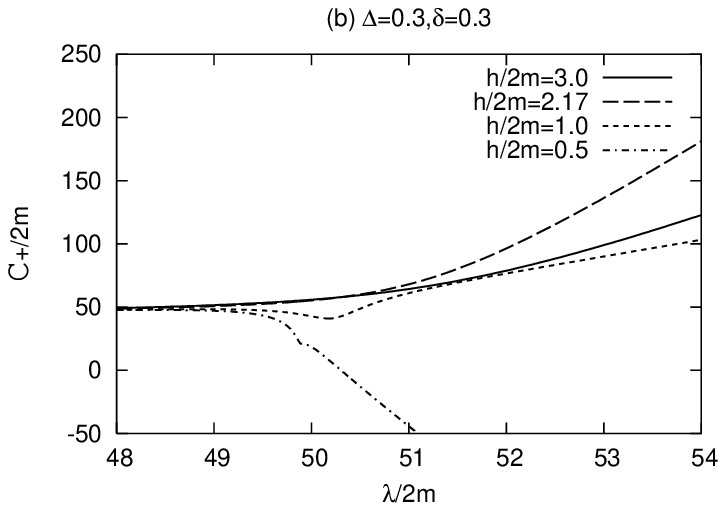}}  \\
  {\includegraphics[width=.41
\textwidth]{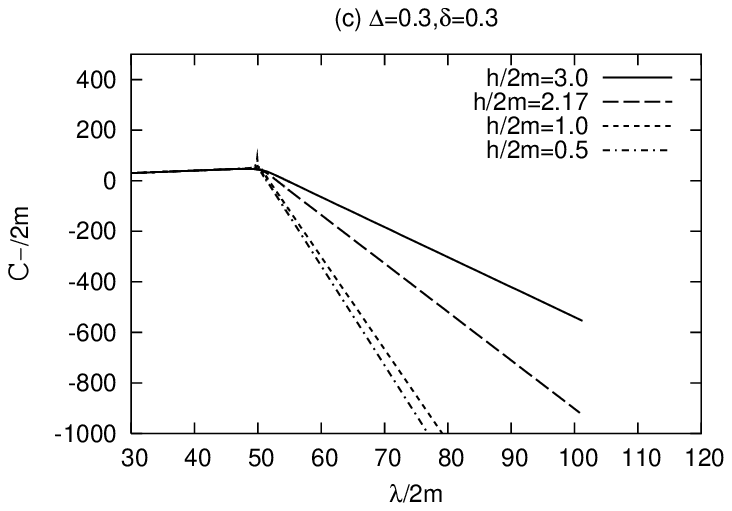}}  &
  {\includegraphics[width=.41
\textwidth]{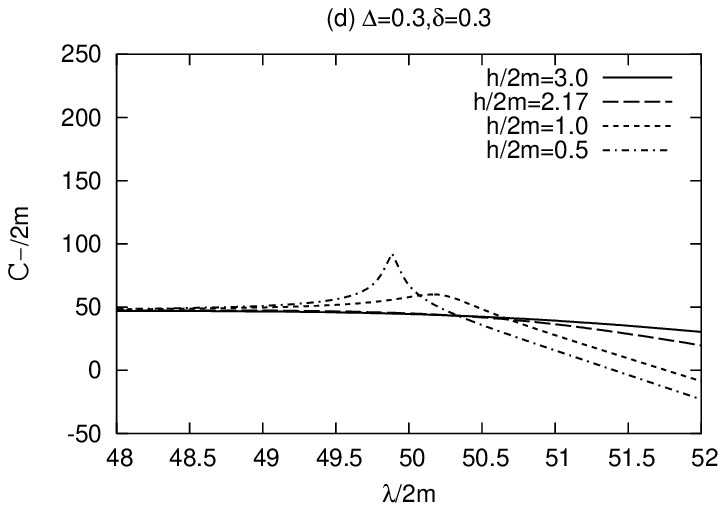}}  
      \label{fig.13}
   \end{tabular}
      \caption{
We show (a) $C_+$ and (c) $C_-$ as
functions of $\lambda$ for several values of
the impact parameters, $h$, as denoted in each figure. 
The model parameters are chosen as
those of the case $\caseNB$, where $(\delta,\Delta)=(0.3,0.3)$.
The figures (b) and (d) are the enlarged parts of
the figures (a) and (c), respectively.
}
  \end{figure}
  \begin{figure}[ht]
 \begin{tabular}{cc}
  {\includegraphics[width=.41\textwidth]{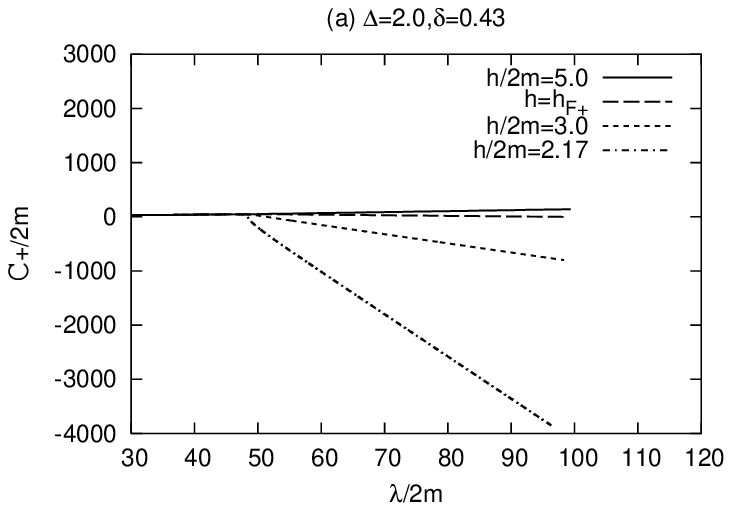}}  &
  {\includegraphics[width=.41\textwidth]{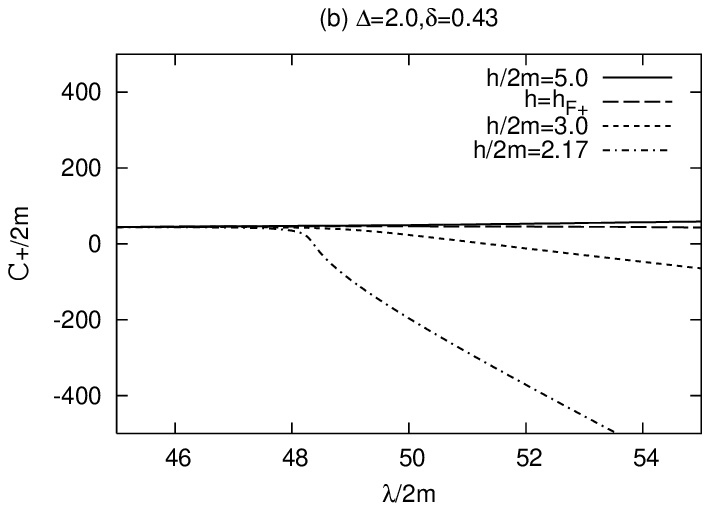}}  \\
  {\includegraphics[width=.41\textwidth]{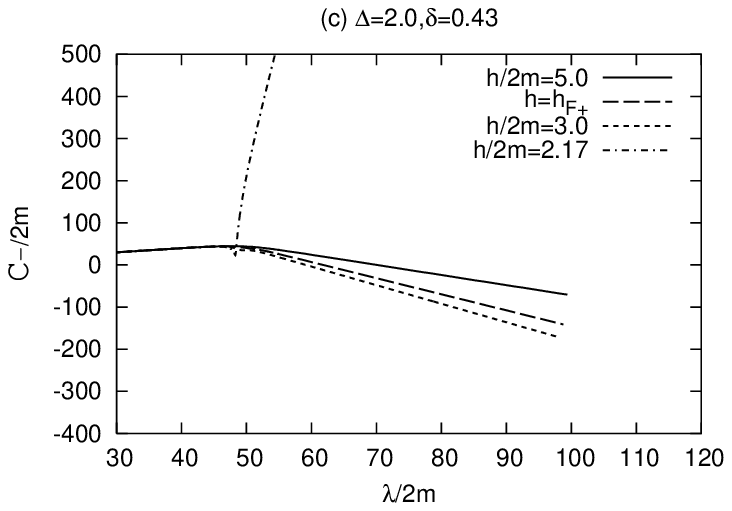}}  &
  {\includegraphics[width=.41\textwidth]{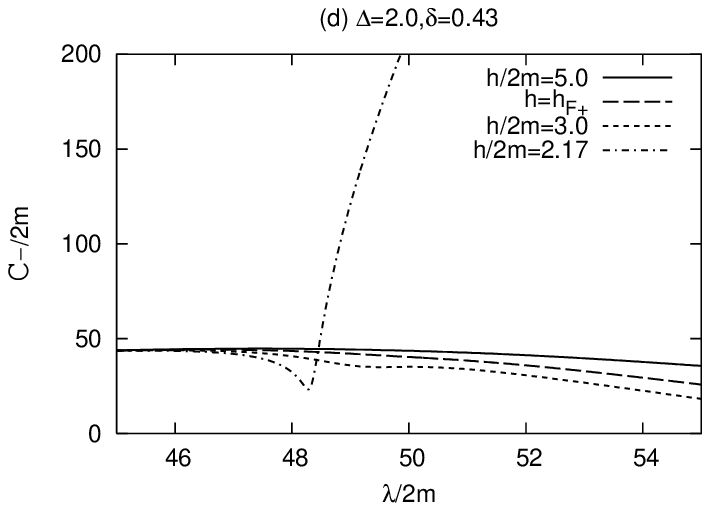}}  
      \label{fig.14}
   \end{tabular}
      \caption{
We show (a) $C_+$ and (c) $C_-$ as
functions of $\lambda$ for several values of
the impact parameters, $h$, as denoted in each figure. 
The model parameters are chosen as
those of the case $\caseI$, where $(\delta,\Delta)=(0.43,2.0)$.
The figures (b) and (d) are the enlarged parts of
the figures (a) and (c), respectively.
}
  \end{figure}

  \begin{figure}[ht]
 \begin{tabular}{cc}
  {\includegraphics[width=.41\textwidth]{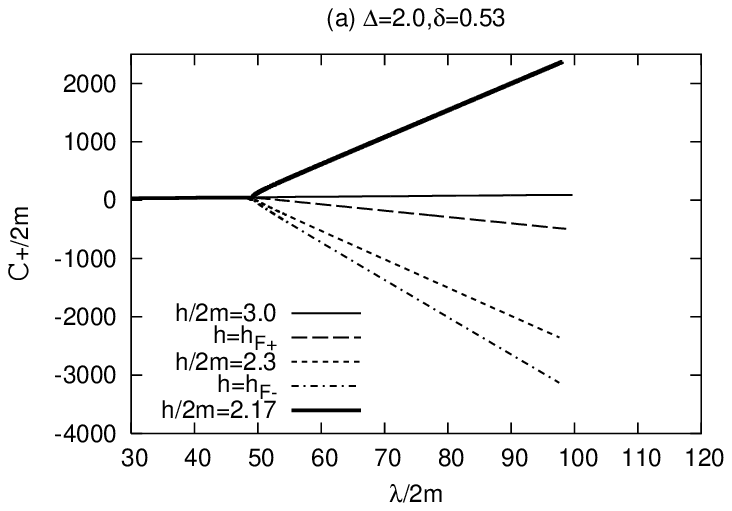}}  &
  {\includegraphics[width=.41\textwidth]{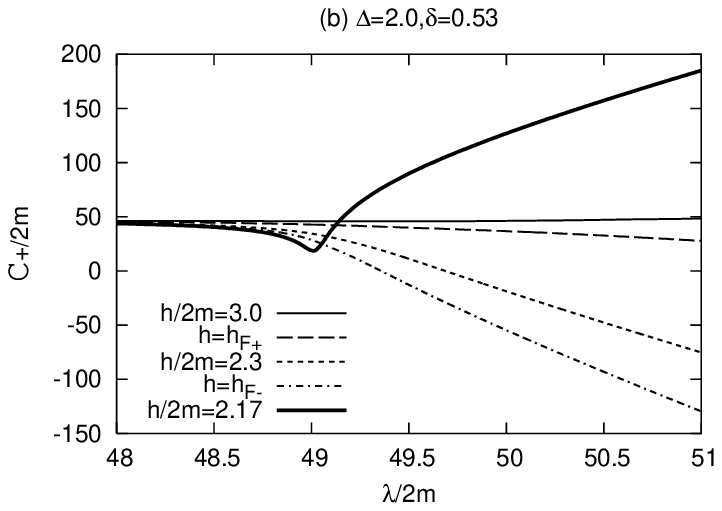}}  \\
  {\includegraphics[width=.41\textwidth]{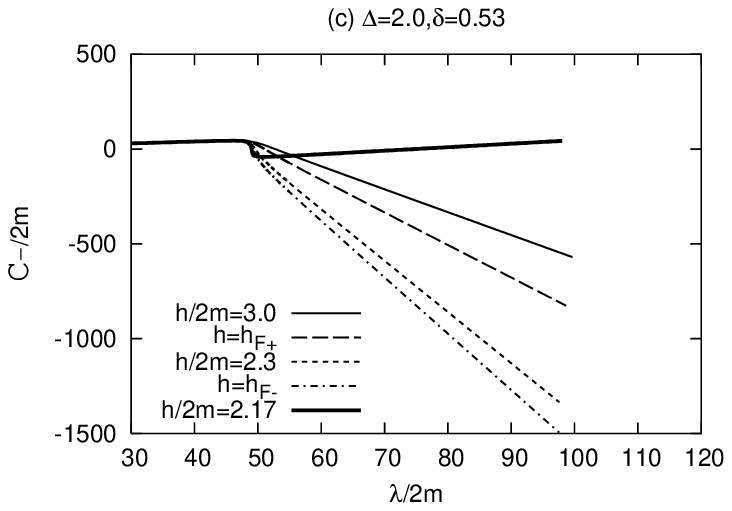}}  &
  {\includegraphics[width=.41\textwidth]{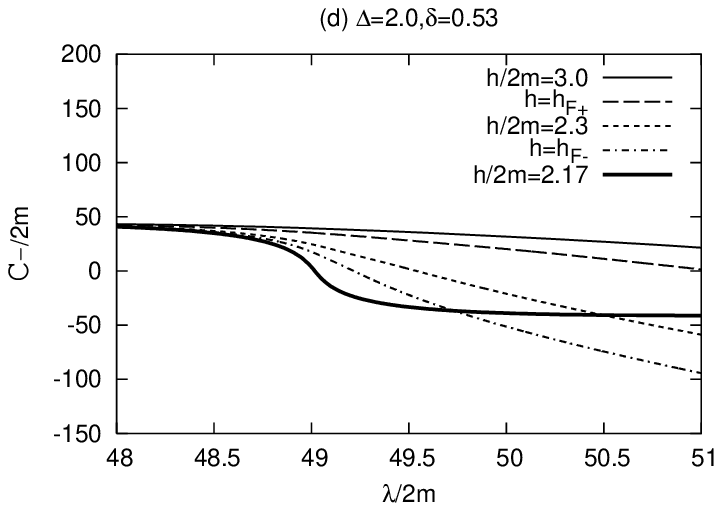}}  
      \label{fig.15}
   \end{tabular}
      \caption{
We show (a) $C_+$ and (c) $C_-$ as
functions of $\lambda$ for several values of
the impact parameters, $h$, as denoted in each figure. 
The model parameters are chosen as
those of the case $\caseII$, where $(\delta,\Delta)=(0.53,2.0)$.
The figures (b) and (d) are the enlarged parts of
the figures (a) and (c), respectively.
}
  \end{figure}

\clearpage
  \begin{figure}[t]
 \begin{tabular}{cc}
  {\includegraphics[width=.41\textwidth]{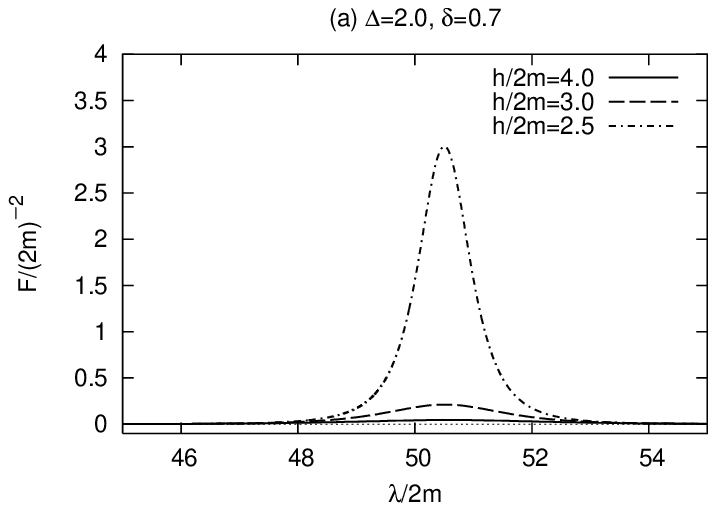}}  &
  {\includegraphics[width=.41\textwidth]{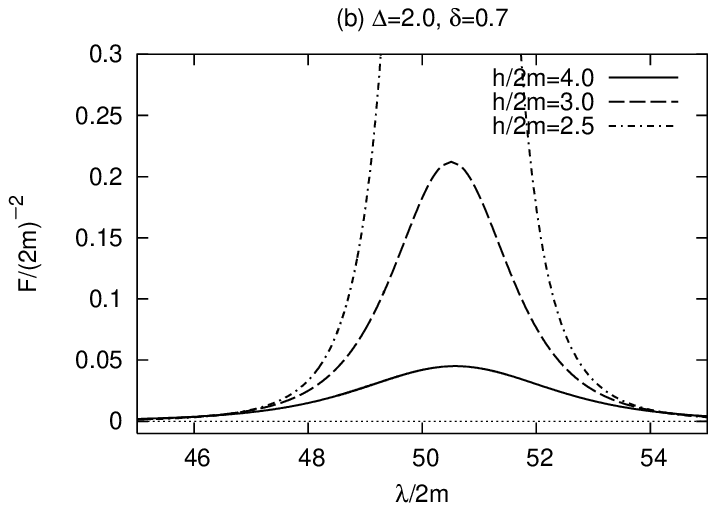}}  \\
  {\includegraphics[width=.41\textwidth]{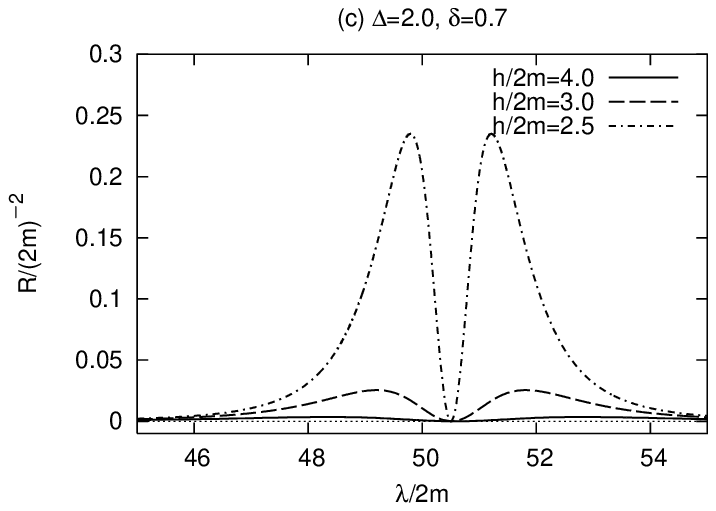}}  &
  {\includegraphics[width=.41\textwidth]{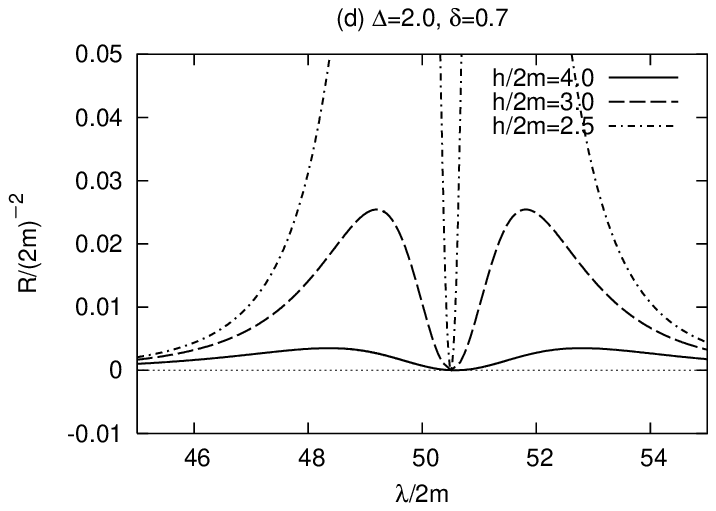}}  
      \label{fig.16}
   \end{tabular}
     \caption{
We show the Weyl and Ricci source-terms, respectively,
in the figures (a) and (c), 
as functions of the affine parameter, $\lambda$,
for several values of the impact parameter, $h$. 
The model parameters are chosen as
those of the case $\caseNA$, where $(\delta,\Delta)=(0.7,2.0)$.
The figures (b) and (d) are the enlarged central parts of
the figures (a) and (c), respectively.
}
  \end{figure}
  \begin{figure}[b]
 \begin{tabular}{cc}
  {\includegraphics[width=.41\textwidth]{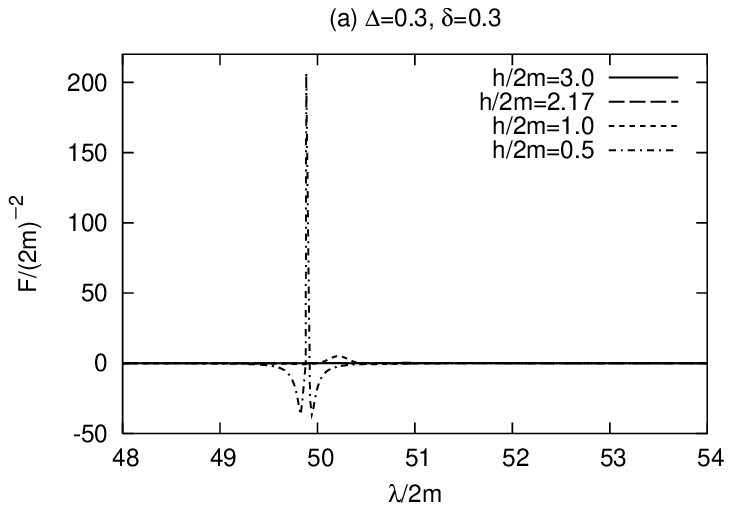}}  &
  {\includegraphics[width=.41\textwidth]{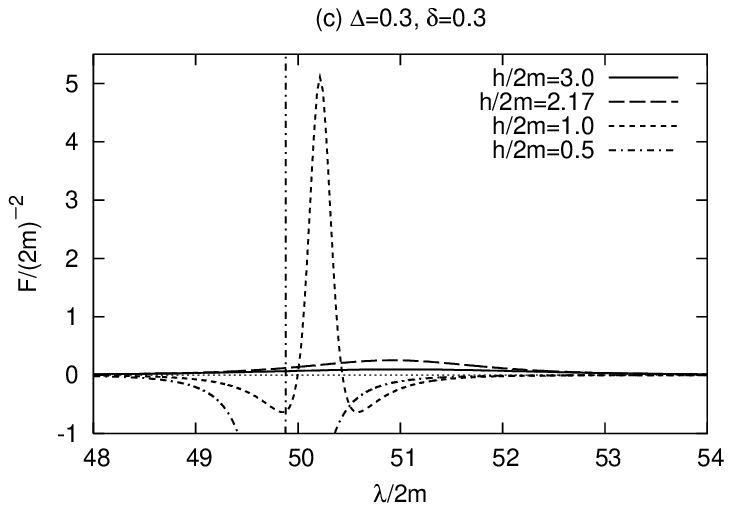}}  \\
  {\includegraphics[width=.41\textwidth]{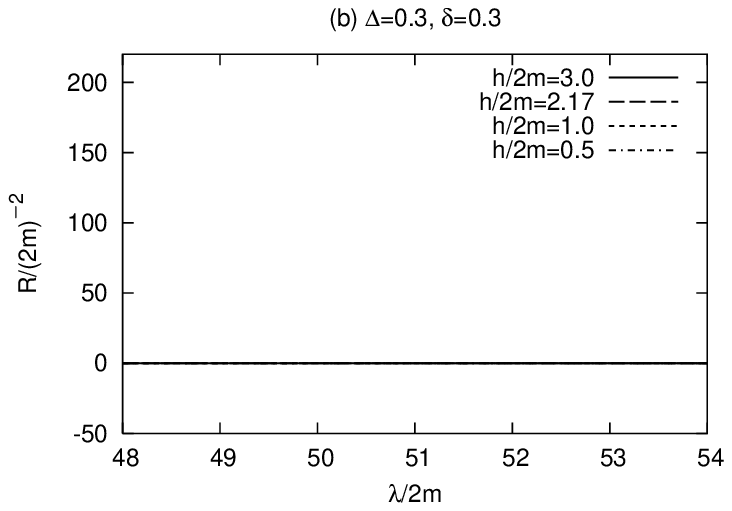}}  &
  {\includegraphics[width=.41\textwidth]{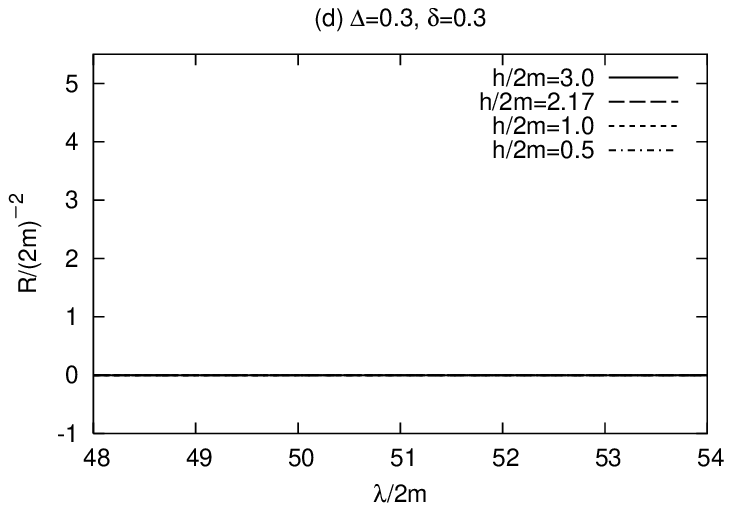}}  
      \label{fig.17} 
   \end{tabular}
     \caption{
We show the Weyl and Ricci source-terms, respectively,
in the figures (a) and (c), 
as functions of the affine parameter, $\lambda$,
for several values of the impact parameter, $h$. 
The model parameters are chosen as
those of the case $\caseNB$: $(\delta,\Delta)=(0.3,0.3)$.
The figures (b) and (d) are the enlarged parts of
the figures (a) and (c), respectively.
The Ricci source-term, ${\cal R}$, is exactly zero
for any value of $h$.
}
  \end{figure}
  \begin{figure}[ht]
 \begin{tabular}{cc}
  {\includegraphics[width=.41\textwidth]{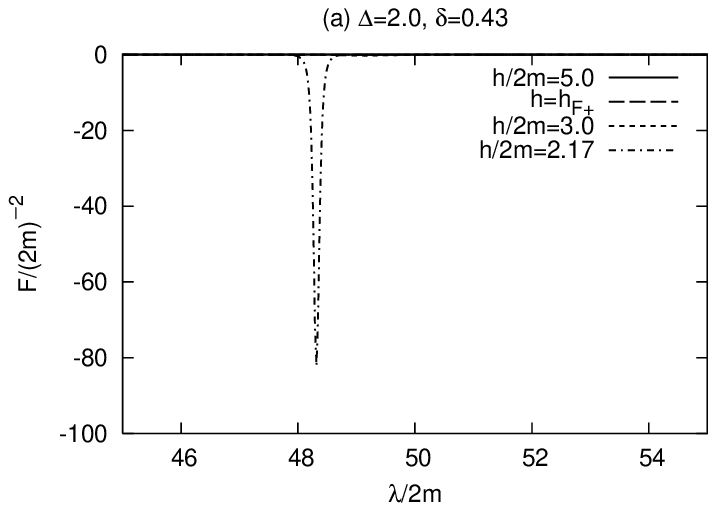}}  &
  {\includegraphics[width=.41\textwidth]{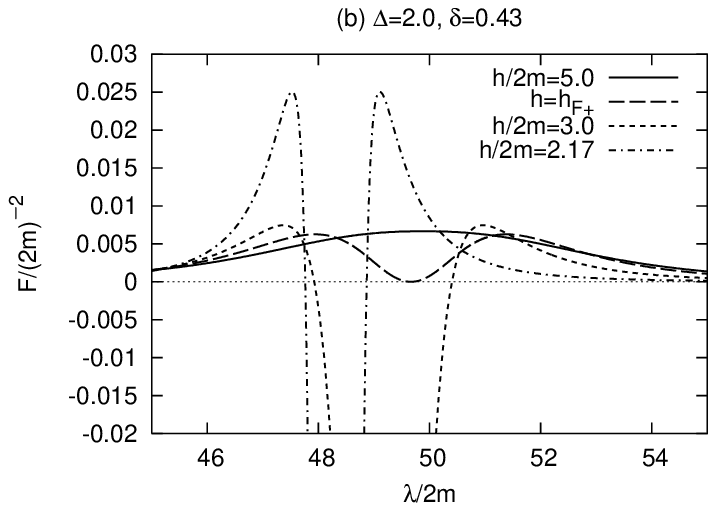}}  \\
  {\includegraphics[width=.41\textwidth]{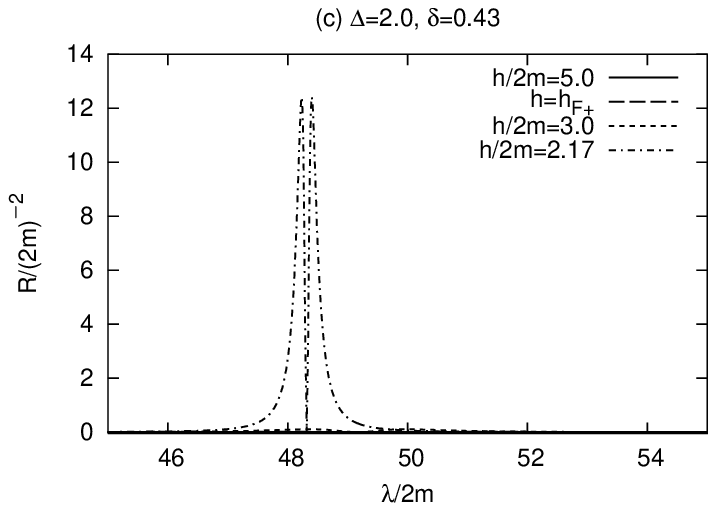}} &
   {\includegraphics[width=.41\textwidth]{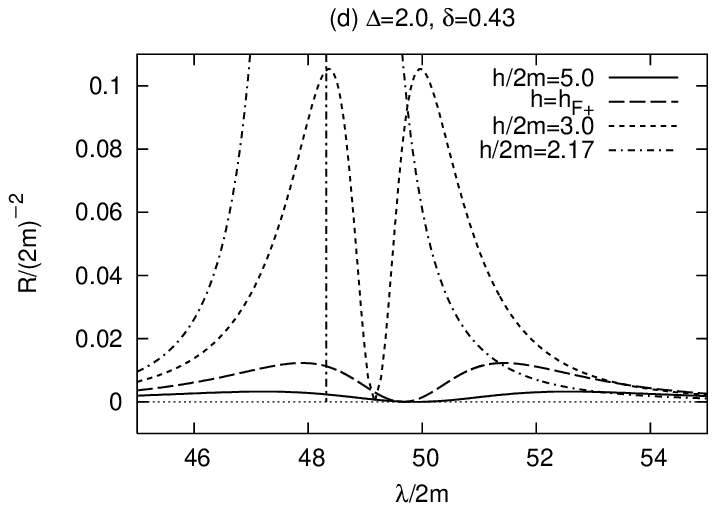}} 
      \label{fig.18}
   \end{tabular}
     \caption{
We show the Weyl and Ricci source-terms, respectively,
in the figures (a) and (c), 
as functions of the affine parameter, $\lambda$,
for several values of the impact parameter, $h$. 
The model parameters are chosen as
those of the case $\caseI$, where $(\delta,\Delta)=(2.0,0.43)$.
The figures (b) and (d) are the enlarged parts of
the figures (a) and (c), respectively.
}
  \end{figure}
  \begin{figure}[ht]
 \begin{tabular}{cc}
  {\includegraphics[width=.41\textwidth]{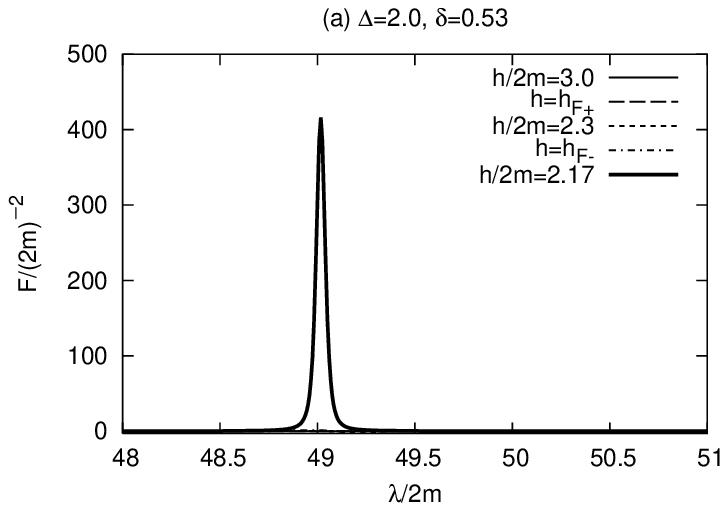}}  &
  {\includegraphics[width=.41\textwidth]{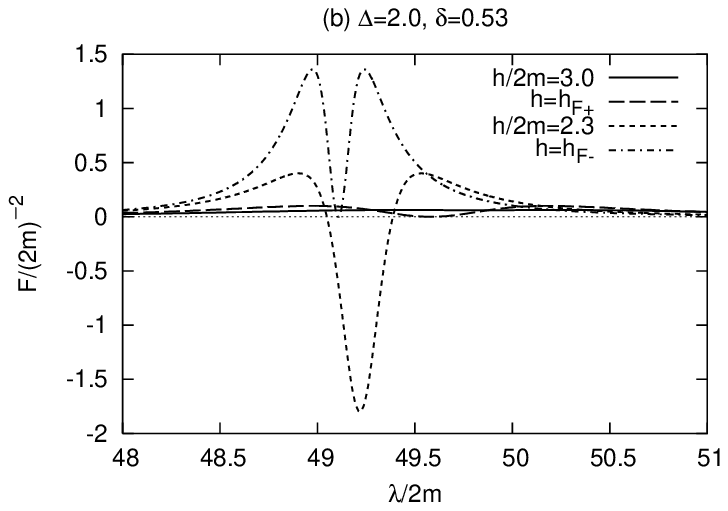}}  \\
  {\includegraphics[width=.41\textwidth]{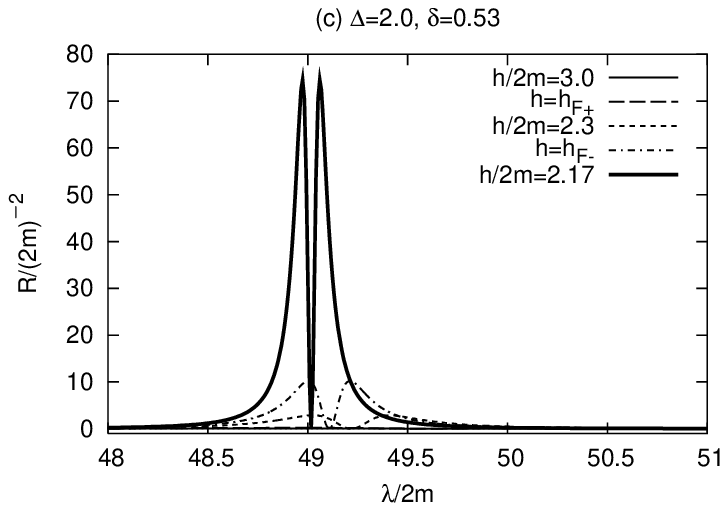}} &
  {\includegraphics[width=.41\textwidth]{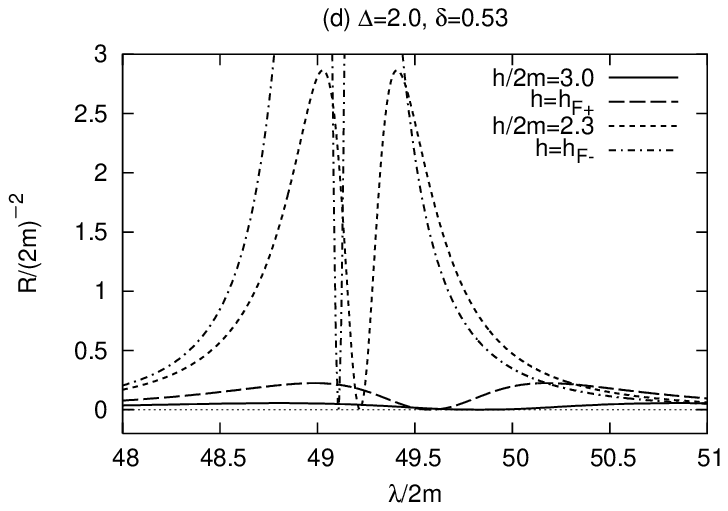}} 
      \label{fig.19}
   \end{tabular}
     \caption{
We show the Weyl and Ricci source-terms, respectively,
in the figures (a) and (c), 
as functions of the affine parameter, $\lambda$,
for several values of the impact parameter, $h$. 
The model parameters are chosen as
those of the case $\caseII$, where $(\delta,\Delta)=(0.53,2.0)$.
The figures (b) and (d) are the enlarged parts of
the figures (a) and (c), respectively.
}
  \end{figure}

\clearpage
\

The oscillatory properties of $C_-$ have been expected by
Dyer's results (see Appendix C), while we find that
$C_+$ can oscillate because its combined source-term,
$-{\cal R}+F$, can be negative not only in the
scalar-tensor-Weyl solution but even in Voorhees's one.
We have tried to find the numerical examples in which
$C_+$ and $C_-$ oscillate.
After finely adjusting the model parameters, $(\delta,\Delta)$,
and the impact parameter, $h$, we find several examples
in which $C_-$ oscillates. As for $C_+$, however, we have failed.
In Figure 20, we show the optical scalars, $C_\pm$,
together with their source-terms for 
the case $\caseII$ with $h/2m=2.1609$.
With Figures 20a and 20b,
we obtain $\epsilon=0.04\times(2m)$ and $\kappa=77/(2m)$
such that $\epsilon\kappa=3.1>1$.
We then estimate the expected number of times of the oscillations
at $\epsilon\kappa/\pi\sim1$, which is in accordance
with the numerical one (see Figure 20f).
As for $C_+$, the analytic result (\ref{uplus}) 
is approximate to the numerical one as follows,
\begin{equation}
\log C_+/2m\sim
\begin{cases}
78\lambda/(2m)+4&~({\rm Analytic})\\
80\lambda/(2m)+6&~({\rm Numerical}).
\end{cases}
\end{equation}
When the combined source-term has plural peaks,
one can superpose the corresponding thin lenses.
  \begin{figure}[ht]
     \begin{tabular}{cc}
    \includegraphics[width=.41\textwidth]{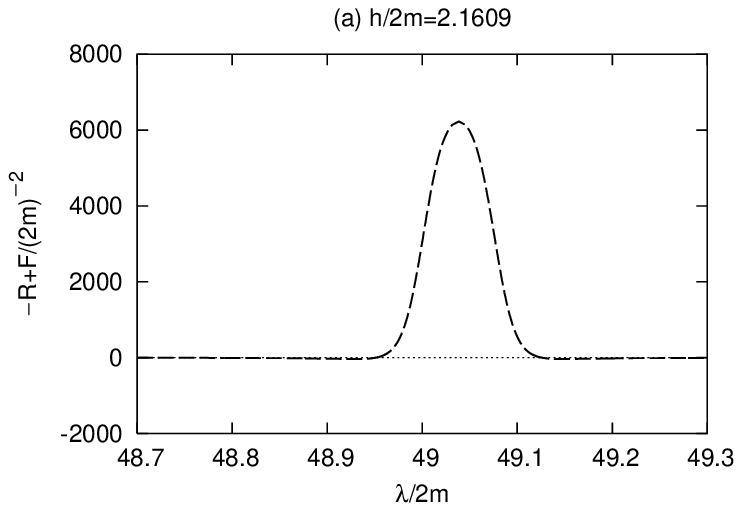} &
    \includegraphics[width=.41\textwidth]{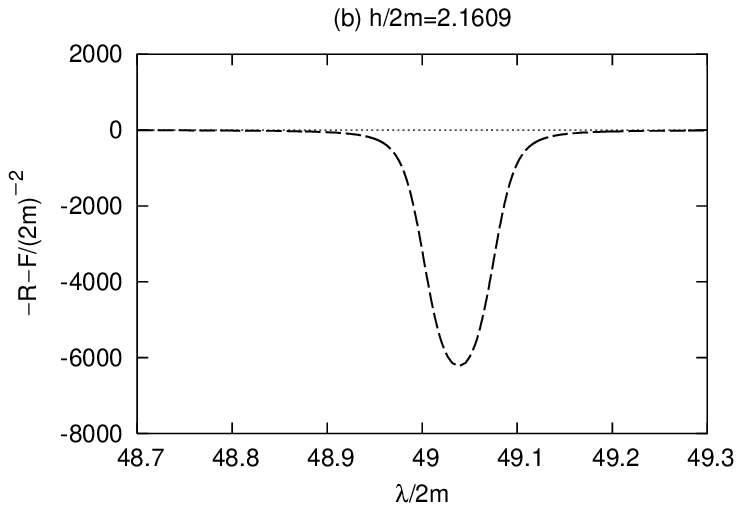} \\
    \includegraphics[width=.41\textwidth]
       {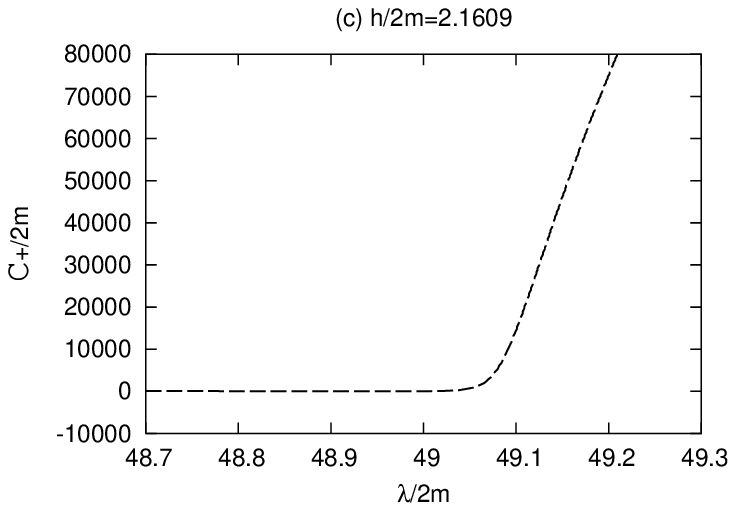} &
     \includegraphics[width=.41\textwidth]
       {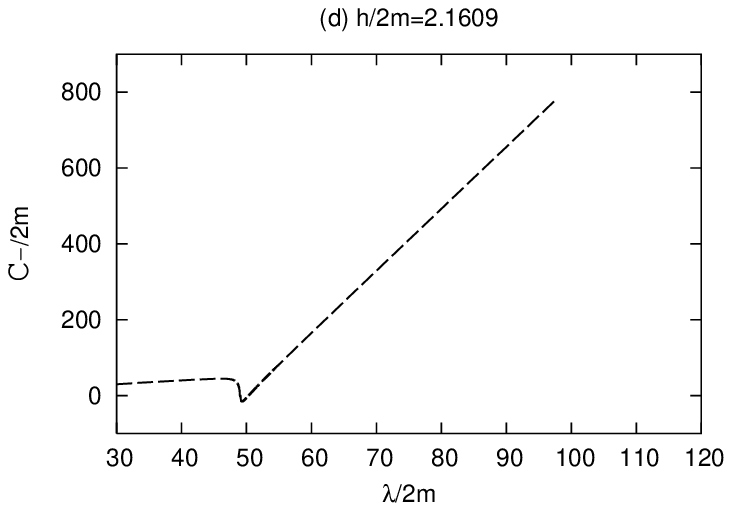} \\
     \includegraphics[width=.41\textwidth]
        {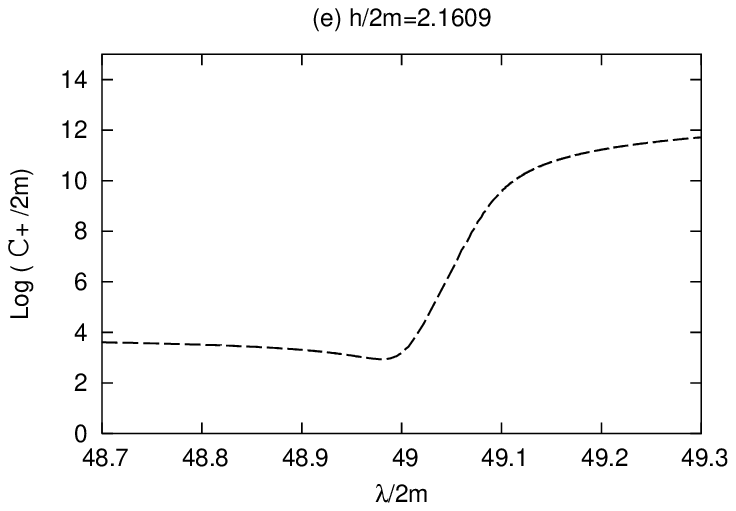} &
     \includegraphics[width=.41\textwidth]
         {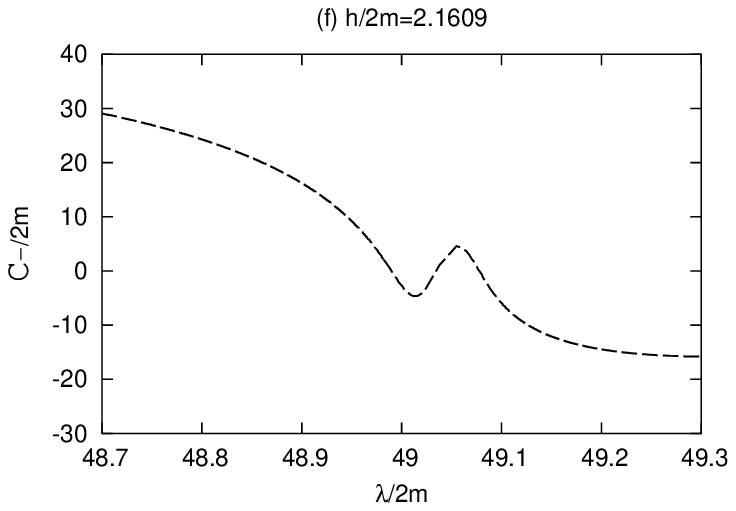}
     \end{tabular}
     \caption{
In the case $\caseII$, where $(\delta,\Delta)=(0.53,2.0)$,
we show (a) $-{\cal R}+F$,~(b) $-{\cal R}-F$,
(c) $C_+$, (d) $C_-$ and (e) $\log C_+$.
The impact parameter, $h$, is finely adjusted
as $h/2m=2.1609$ such that $C_-$ oscillates.
The figure (f) is the enlarged part of the figure (d). 
}
      \label{fig.20}
  \end{figure}

\

Now we summarize the qualitative properties 
of $C_+$ and $C_-$ along the light path 
for each region. 

\

\begin{description}
\item[(${{\bf N}_{\bf A}}$)]~~In the region ${{\rm N}_{\rm A}}$,
the Weyl source-term, $F$, is always dominant over the Ricci
source-term, ${\cal R}$, and moreover is positive-definite.
When the impact parameter is larger enough than $m$,
the inequality, $\epsilon\kappa<<1$, holds, and
the bending angle of the graph monotonously increases
as the impact parameter decreases.
Accordingly, there is a single caustic point, $h=h_a$, near 
the Einstein radius, $h_E$, such that $C_-$ vanishes at
the final time, $\lambda=\lambda_f$. 
For the smaller impact parameter, 
the inequality, $\epsilon\kappa>1$, holds, and
$C_-$ can oscillate.

\item[(${{\bf N}_{\bf B}}$)]~In the region ${{\rm N}_{\rm B}}$,
the Weyl source-term, $F$, is always dominant over the Ricci
source-term, ${\cal R}$.
When the impact parameter is large enough,
the properties of $C_+$ and $C_-$ are similar to those
in the region ${{\rm N}_{\rm A}}$.
For the smaller impact parameter, 
the Weyl source-term can be negative
such that $C_+$ can decrease, while $C_-$ can increase.
Accordingly, there is a new caustic point, $h=h_b$,
with $h_b=3m\sim 6m$ such that $C_+$ vanishes at the final time.
The intriguing thing for $h<h_b$ is that $C_+$ has negative values
with the very large amplitude, while $C_-$ is moderately positive. 
The case of $C_-$ with $h/2m=0.5$ may be an example
of the superposition of triple lenses (see Figure 17a). 

\item[(~I~)]~In the region I, 
the Weyl source-term, $F$, is mostly dominant over the Ricci
source-term, ${\cal R}$, and the Ricci-dominant cases
are found only in the very narrow range of the
impact parameter around $h_{F+}$.
When $h_{F+}<h_E$, a first caustic point exists near $h_E$,
and a second caustic points, $h=h_b$, appears near $h_{F+}$
such that $C_{+}$ vanishes at the final time.
It should be noted that
the inequality, $h_{F+}< h_E$, is easily
violated in our numerical studies with $x_i=100\delta$ 
(see Figure 23). In these cases, the caustic point, $h=h_b$,
becomes the first one.
Let $C_+$ and $C_-$ at the final time, $\lambda=\lambda_f$, 
be functions of the impact parameter, $h$. 
For $h<h_b$, $C_+$ becomes negative with a larger amplitude
for smaller $h$, while $C_-$ changes its sign again 
and becomes positively larger for smaller $h$. 
Accordingly, a third caustic point can exist
such that $C_{-}$ vanishes at the final time.

\item[(II)]~In the region II, 
the Weyl source-term, $F$, is mostly dominant over the Ricci
source-term, ${\cal R}$, and the Ricci-dominant cases
are found only in the very narrow ranges of the
impact parameter around $h_{F+}$ and $h_{F-}$.
We find the very complicated structure of caustic points.
For example, in the case \caseII, we find four caustic points,
$h_a$, $h_b$, $h_c$ and $h_d$, where $h_a>h_b>h_c>h_d$ holds.
The first caustic point, $h_a$, appears near the Einstein radius, $h_E$,
such that $C_-$ vanishes at the final time.
The second and third ones, $h_b$ and $h_c$,
appear near $h_{F+}$ and $h_{F-}$, respectively,
such that $C_+$ vanishes at the final time.
Finally, the fourth one is associated with $C_-$
such that $C_-$ vanishes at the final time.
\end{description}

\

We have defined a conformally invariant variable, $C$, by
$C=(C_+-C_-)/C_+$, which can be interpreted as
the image distortion rate. In Figure 21, we show
$C$ at the final time as a function of the impact parameter, $h$. 
As is noted above, one of the most intriguing
properties of the graph is
the appearance of the caustic points, and we find 
the following four cases
of the number, $N_+$, of the caustic 
points associated with $C_+$:

\begin{center}
(a)~~$N_+=0$,~~~~~(b)~~$N_+=1$,~~~~~(c)~~$N_+=2$,
~~~~~(d)~~$N_+\neq0$ and $h_E<h_{F+}$.
\end{center}

\noindent
Now we should remind ourselves that $h_E$ is an extrinsic
scale including the initial distance to the source, 
$R_i=R_x(x_i)$.
It is naturally expected that the cases of (d) are distributed
to (b) and (c).
As such an example, we investigate the cases of 
$(\delta,\Delta)=(2.0,0.3)$ in the region I and
$(\delta,\Delta)=(4.0,0.6)$ in the region II.
In both cases, we have $h_{F+}/2m>h_E/2m\sim7$ for $x_i=100\delta$
and $h_{F+}/2m<h_E/2m\sim14$ for $x_i=400\delta$.
We show the image distortion rate, $C$, as a function
of the impact parameter, $h$, in Figure 22. 
The graphs show the ``anomalies'' for the cases of $x_i=100\delta$
in the sense that the first caustic point, $C=1$, near
the Einstein radius, $h_E$, disappears.
That is, a naked caustic point associated with $C_+$ appears
near $h_{F+}$. For the cases of $x_i=400\delta$,
this anomalous caustic point is hidden by
the Einstein radius, $h_E$, such that the first caustic point
appears near $h_E$ as the one associated with $C_-$.
We classify the parameter space, $(\delta,\Delta)$, 
on the basis of the above conditions, (a)$\sim$(d),
as is depicted in Figure 23.
Again, we find a close relationship 
between the above new classification and the previous one.

\

We summarize the relationship between our numerical results
and the classification of the parameter space, $(\delta,\Delta)$, 
in Table I.

  \begin{figure}[t]
 \begin{tabular}{cc}
  {\includegraphics[width=.45\textwidth]{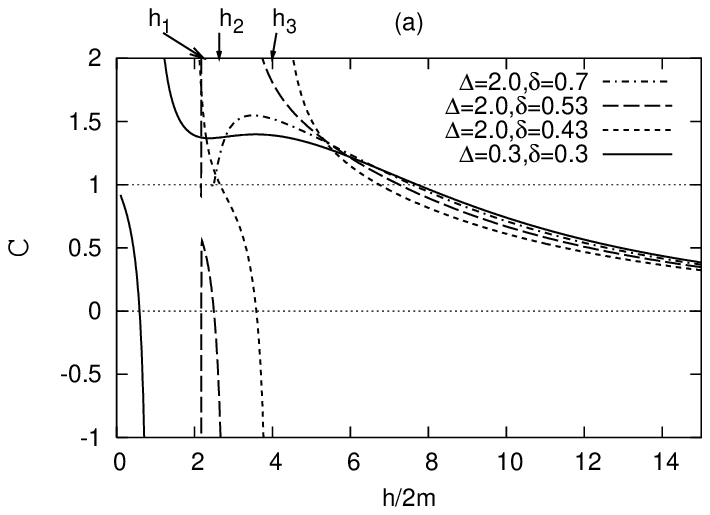}}  &
  {\includegraphics[width=.45\textwidth]{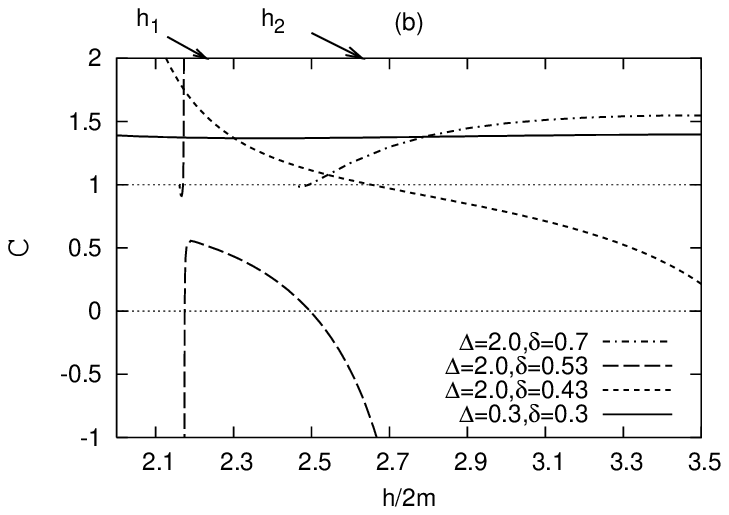}}  
      \label{fig.21}
   \end{tabular}
   \caption{
In the figure (a), the image distortion rate, $C$, is shown 
as a function
of the impact parameter, $h$.
The model parameters are chosen to be the standard ones.
The figure (b) is the enlarged part of the figure (a).
    }
  \end{figure}

  \begin{figure}[ht]
 \begin{tabular}{cc}
  {\includegraphics[width=.45\textwidth]{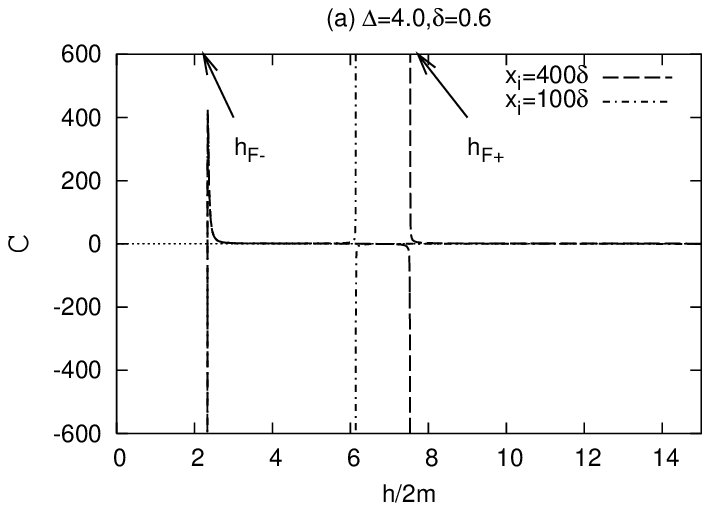}}  &
  {\includegraphics[width=.45\textwidth]{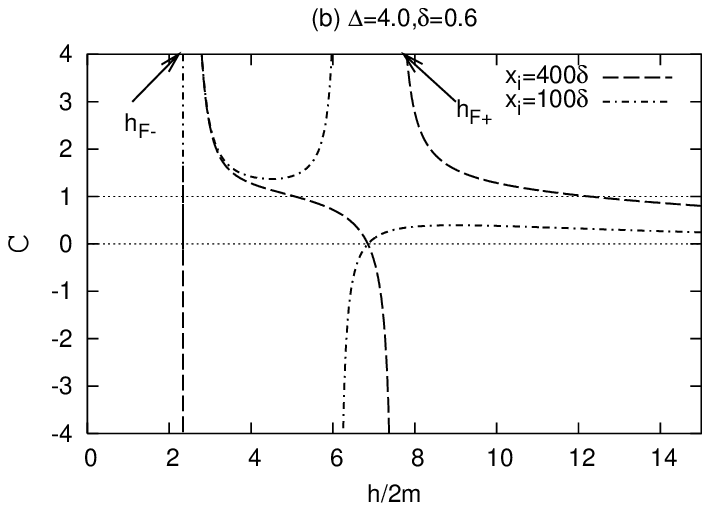}}  \\
  {\includegraphics[width=.45\textwidth]{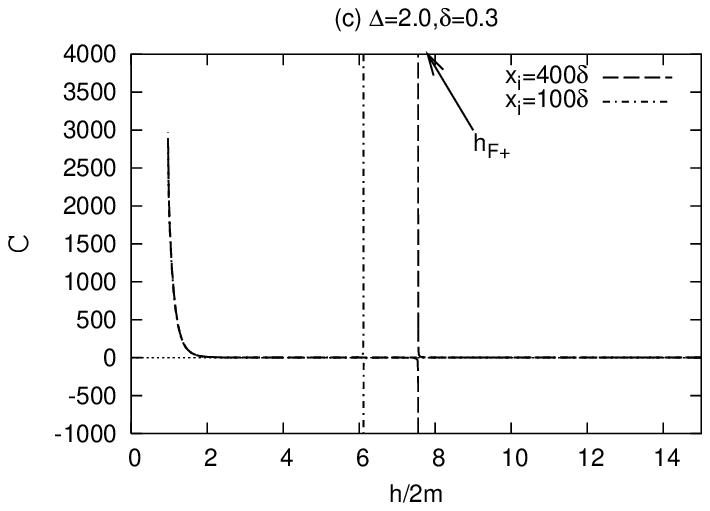}}  &
  {\includegraphics[width=.45\textwidth]{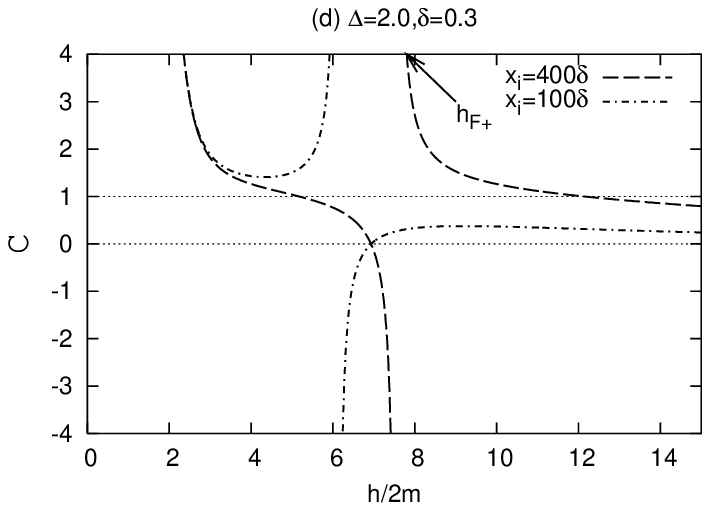}}  
      \label{fig.22}
   \end{tabular}
\caption{ 
We show the image distortion rate, $C$, as a function
of the impact parameter, $h$, for two cases: 
(a) $(\delta,\Delta)=(0.6,4.0)$ in the region II and
(c) $(\delta,\Delta)=(0.3,2.0)$ in the region I.
The enlarged figures (b) and (d) correspond to the figures (a)
and (c), respectively.
}
  \end{figure}
  \begin{figure}[ht]
 \begin{tabular}{cc}
  \centerline{\rotatebox{-90}{{\resizebox{50mm}{!}{\includegraphics{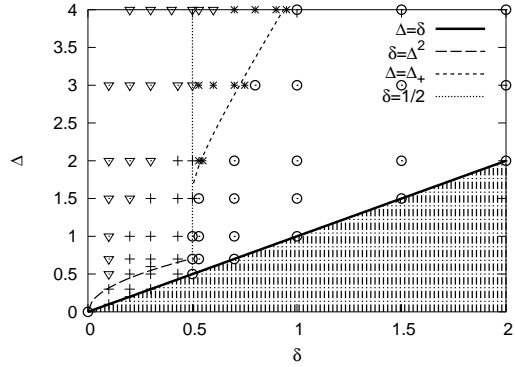}}}}}  
      \label{fig.23}
   \end{tabular}
\caption{ 
We classify
the parameter space, $(\delta,\Delta)$, into the four groups, 
(a)$\sim$(d), on the basis of the number of the caustic points, $N_+$, 
associated with $C_+$, as is explained in the text:
(a) an open circle,~(b)~$+$,~(c)~$*$~and~(d) a reversed triangle. 
}
  \end{figure}
\begin{table}[ht]
\caption{The classification of the parameter space, $(\delta,\Delta)$}
\begin{center}
   \begin{tabular}{c|ccccc}  \hline \hline
Classification of $F(x_o)$    &   N$_A$       & N$_B$  
& I   & II   & Fig. 3   \\ \hline
Reflecting region             & mostly blank  & $+$    
& $+$ & $+$  & Fig. 7 \\ \hline
Extremes of the deflection angle & mostly blank  & $+$    
& $+$ & $\ast$ & Fig. 10 \\ \hline 
Negative deflection angle & blank  & $\bullet$    
& $\bullet$ & mostly $\bullet$ & Fig. 11 \\ \hline 
Caustic points associated with $C+$ & mostly $\circ$  & $+$    
& $+$ & $\ast$ & Fig. 23 \\ \hline 
    \end{tabular}
\end{center}
\end{table}
\clearpage
\

\section{A closing remark}

\

In this paper, we have investigated several properties
of null geodesics and null geodesic congruences with 
the aim of finding the crucial, experimentally
detectable difference
between the theory of general relativity 
and scalar-tensor theories of gravity. 
Our studies are very restricted ones because
they are performed only for specific null geodesics (ones on
the equatorial plane, $y=0$) with a specific 
solution (the scalar-tensor-Weyl solution).
These specific null geodesics may be unstable
when there are additional small fluctuations,
especially those in the $y$-direction.
One may doubt the realities of the specific solution
as a mathematical model of the spacetime.
Indeed we cannot guarantee its realities.
Nevertheless, we would like to claim that
we have succeeded in discovering the new world,
namely the region II in the classified parameter space. 
The lensing properties found
in this region are never seen in the other regions, 
${{\rm N}_{\rm A}}$, ${{\rm N}_{\rm B}}$ and I. 
Unfortunately, the border between the region II (the new world)
and the region ${{\rm N}_{\rm A}}$ (the Einstein-like world)
is rather vague, and moreover, the most precious
things ($h_{F+}$ and $h_{F-}$) may be hidden by
the Einstein radius, $h_E$. 
But one will go there: $someday,~somewhere,~somehow$.\cite{15}

\
\textit{Acknowledgements} We would like to thank 
Ken-ichi Oohara and Ryoichi Nishi for helpful comments.
\appendix
\section{The Ernst-like equations in scalar-tensor theories}

\

In the stationary, axisymmetric spacetime, 
there exist two Killing vectors, $\boldsymbol{\xi}=\partial_t$ and 
$\boldsymbol{\eta}=\partial_{\phi}$. 
It is shown that the energy-momentum tensor,
$T^{\alpha\beta}$, of the Maxwell fields satisfies
the following relations\cite{14},
\begin{equation}
\xi^{\alpha}T_{\alpha}^{\ [\beta}\xi^{\gamma}\eta^{\delta]}=
\eta^{\alpha}T_{\alpha}^{\ [\beta}\xi^{\gamma}\eta^{\delta]}=
0.
\label{a01}
\end{equation}
One finds that
the similar relations hold
for $S_{\mu\nu}\equiv
2\partial_{\mu}\varphi\partial_{\nu}\varphi$, 
accordingly, with Theorem 7.1.1 in Ref.13, 
the metric of the Einstein-Maxwell system with 
a scalar field, $\varphi$, can be reduced to the following form,
\begin{equation}
ds^2=-\e^{2\psi}(dt-\omega d\phi)^2 +
\e^{-2\psi}\left[\e^{2\gamma}(d\rho^2+dz^2)+\rho^2d\phi^2\right],
\label{a1}
\end{equation}
where the metric functions, $\psi,\gamma$ and $\omega$, 
are functions of $x^1\equiv\rho$ and $x^2\equiv z$.

After long and complicated calculations, one finds that 
the field equations are reduced to
the following set of equations, 
\begin{subequations}
\begin{eqnarray}
\e^{2\psi} & = & \frac{1}{2}\left({\EP}+\bar{\EP}\right)+\Phi\bar{\Phi}, 
\label{a7} \\
\e^{2\psi}\nabla^2{\EP} & = &
(\nabla\EP)\cdot\left[(\nabla\EP)+2\bar{\Phi}\nabla\Phi\right], 
\label{a8} \\
e^{2\psi}\nabla^2{\Phi} & = &
(\nabla\Phi)\cdot\left[(\nabla\EP)+2\bar{\Phi}\nabla\Phi\right], 
\label{a9} \\
\gamma_{,1}&=&
\rho\left[(\psi_{,1})^2-(\psi_{,2})^2\right]
-\frac{1}{4\rho}\left[(\omega_{,1})^2-(\omega_{,2})^2\right]
\e^{4\psi} \nonumber \\
& & -\rho(\Phi_{,1}\bar{\Phi}_{,1}-\Phi_{,2}\bar{\Phi}_{,2})\e^{-2\psi} 
    +\rho\left[(\varphi_{,1})^2-(\varphi_{,2})^2\right],
\label{a10} \\
\gamma_{,2}&=&
2\rho\psi_{,1}\psi_{,2}
-\frac{1}{2\rho}\omega_{,1}\omega_{,2}\e^{4\psi}
-\rho(\Phi_{,1}\bar{\Phi}_{,2}+\Phi_{,2}\bar{\Phi}_{,1})\e^{-2\psi}
+2\rho\varphi_{,1}\varphi_{,2},
\label{a11} \\
\nabla^2\varphi & = & 0, 
\label{a12}
\end{eqnarray}
\end{subequations}
where, for any functions, $f$ and $h$,
\begin{equation}
\nabla^2 f\equiv\frac{1}{\rho}\frac{\partial}{\partial\rho}
\left(\rho\frac{\partial f}{\partial\rho}\right)
+ \frac{\partial^2 f}{\partial z^2},~~~
\nabla f\cdot\nabla h\equiv
\frac{\partial f}{\partial \rho}\frac{\partial h}{\partial \rho}+
\frac{\partial f}{\partial z}\frac{\partial h}{\partial z},~~~
f_{,i}\equiv\frac{\partial f}{\partial x^i}.
\end{equation}
In Ref.14,
there are explicit forms of the metric function, $\omega$, and
the Maxwell fields in terms of the complex potentials, 
${\cal E}$ and $\Phi$. 
One notes that the equations (A.3a)$\sim$(A.3c) are the same as those in
general relativity and are referred to as 
the Ernst equations of the Einstein-Maxwell system. 
The remaining equations contain the scalar field contributions,
and, as for the metric functions, 
effects of the scalar field  appear only
in $\gamma$.

\

\section{The scalar-tensor-Weyl solution}

\

In the case of static, axisymmetric
vacuum solutions,
one has $\omega=\Phi=0$, and
the field equations (\ref{a7})$\sim$(\ref{a12})
become the following set of equations,
\begin{subequations}
\begin{eqnarray}
\e^{2\psi} &=& \EP,\\
\EP\nabla^2{\EP} & = & (\nabla\EP)\cdot(\nabla\EP), 
\label{b2} \\
\gamma_{,1}&=&
\rho\left[(\psi_{,1})^2-(\psi_{,2})^2\right]
+\rho\left[(\varphi_{,1})^2-(\varphi_{,2})^2\right],
\label{b3} \\
\gamma_{,2}&=&
2\rho\psi_{,1}\psi_{,2}
+2\rho\varphi_{,1}\varphi_{,2},
\label{b4} \\
\nabla^2\varphi & = & 0,
\label{b5}
\end{eqnarray}
\end{subequations}
where ${\EP}$ is a real function.
One finds that the equation for $\psi$ is reduced to
\begin{equation}
\nabla^2\psi = 0.
\label{b7}
\end{equation}
That is, $\varphi$ and $\psi$ are harmonic functions.

We introduce 
$oblate$
and 
$prolate$
coordinates, $(x,y)$, 
defined by
\begin{subequations}
\begin{eqnarray}
\rho & =& \sigma \sqrt{(x^2+\epsilon)(1-y^2)} , 
\label{b8} \\
z & = & \sigma xy ,
\label{b9}
\end{eqnarray}
\end{subequations}
where $\sigma$ is a positive constant, and $\epsilon=\pm 1$.
The cases, $\epsilon=1$ and $\epsilon=-1$, are referred to as
$oblate$ and $prolate$, respectively. 
The equation (\ref{b7}) is then reduced to
\begin{equation}
\frac{\partial}{\partial
  x}\left[(x^2+\epsilon)\frac{\partial\psi}{\partial x}\right]
+\frac{\partial}{\partial
  y}\left[(1-y^2)\frac{\partial\psi}{\partial y}\right] =0 .
\label{b10}
\end{equation}
A similar equation holds for $\varphi$.

The simplest prolate solution for $\psi$ is given by
\begin{equation}
\psi=\frac{\delta}{2}\ln\left(\frac{x-1}{x+1}\right),
\label{b11}
\end{equation}
where $\delta$ is an integration constant.
A similar solution for $\varphi$ is given by
\begin{equation}
\varphi = \varphi_0 + \frac{d}{2}\ln\left(\frac{x-1}{x+1}\right) ,
\label{b13}
\end{equation}
where $\varphi_0$ and $d$ are integration constants.
Then the corresponding metric function, $\gamma$, becomes
\begin{equation}
\e^{2\gamma}=\left(\frac{x^2-1}{x^2-y^2}\right)^{{\Delta}^2} ,
\label{b14}
\end{equation}
where 
\begin{equation}
{\Delta}^2\equiv \delta^2 +d^2 .
\label{b15}
\end{equation}
For completeness, we give an explicit form of the metric:
\begin{eqnarray}
\eqalign{
ds^2 &= -\left(\frac{x-1}{x+1}\right)^{\delta} dt^2 
+\sigma^2 \left(\frac{x-1}{x+1}\right)^{-\delta} \times\cr
& \left[\left(\frac{x^2-1}{x^2-y^2}\right)^{\Delta^2}
(x^2-y^2)\left(\frac{dx^2}{x^2-1}+\frac{dy^2}{1-y^2}\right)
+(x^2-1)(1-y^2) d\phi^2\right].\cr}
\label{b16}
\end{eqnarray}
One notes that, though $\psi$ and $\varphi$ are independent of $y$, 
$\gamma$ depends on both $x$ and $y$.

\section{Analytic solutions to the optical scalar equations}
\label{ap10}

\subsection{Static, spherically symmetric spacetime}

\

We summarize the analytic results obtained by Dyer\cite{11}.
A metric of
the static, spherically symmetric spacetime is given by
\begin{equation}
ds^2=-\e^{2C}dt^2+\e^{2A}dr^2+\e^{2B}d\Omega^2,
\label{gap1}
\end{equation}
where $A$, $B$ and $C$ are functions of $r$.
A null vector, $k^\mu$, which is tangent to the null geodesic
in this spacetime, is obtained as
\begin{equation}
\eqalign{
&k^0=\e^{-2C},~~
k^2=0,~~k^3=h\e^{-2B},\cr
&k^1=\pm\e^{-(A+C)}\sqrt{1-h^2\e^{2(C-B)}},\cr}
\label{gap2}
\end{equation}
where a constant, $h$, is an impact parameter, and
we assume that the geodesic is on the equatorial plane,
$\theta=\pi/2$, without loss of generality. 
The complex null vector, $t^\mu$, becomes
\begin{equation}
\eqalign{
&t^0=\frac{\e^{-C}}{2\sqrt{2}}\left[\frac{2}{h}S_+S_-\e^{B}
+H(S_+^2+S_-^2)\right],~~t^2=\frac{i}{\sqrt{2}}\e^{-B},\cr
&t^1=\frac{\e^{-A}}{2\sqrt{2}}\left[\frac{1}{h}(S_+^2+S_-^2)\e^{B}
+2HS_+S_-\right],~~t^3=\frac{Hh}{\sqrt{2}}\e^{-2B},\cr
&S_\pm=\sqrt{\e^{-C}\pm h\e^{-B}},~~
H=-\frac{1}{h}\int^r\frac{B'}{S_+S_-}\e^{B-C}dr,
}
\label{gap3}
\end{equation}
where a prime denotes a differentiation with respect to $r$.
Then the Ricci and Weyl source-terms, ${\cal R}$ and $F$, 
in (\ref{ge6a}) and (\ref{ge6b}) are evaluated as
\begin{equation}
\eqalign{
&-{\cal R}+F=\e^{-2(A+C)}\left(B''+(B')^2-B'C'-B'A'\right)\cr
&~~~~~~~~~~~-h^2\e^{-2(A+B)}\left(C''+(C')^2-C'B'-C'A'\right),
\cr
&-{\cal R}-F=\e^{-2(A+C)}\left(B''+(B')^2-B'C'-B'A'\right)\cr
&~~~~~~~~~~~-h^2\e^{-2(A+B)}\left(\e^{2(A-B)}+B''-A'B'\right).
\cr}
\label{gap4}
\end{equation}
Since the Weyl source-term, $F$, is real, 
one can let the shear, $\sigma$, be real without
loss of generality.
Dyer has introduced the new optical scalars, $C_\pm$, defined by
\begin{equation}
\frac{d}{d\lambda}\ln C_\pm=\theta\pm\sigma,
\label{gap5}
\end{equation}
and has obtained the following equations,
\begin{equation}
\frac{d^2C_\pm}{d\lambda^2}=(-{\cal R}\pm F)C_\pm.
\label{gap6}
\end{equation}
General solutions to (\ref{gap6}) are given as
\begin{equation}
\eqalign{
&C_+=C_+^0\sqrt{\e^{2B}-h^2\e^{2C}}\left\{\int^{r}
\frac{\e^{A+B+C}}{\left[\e^{2B}-h^2\e^{2C}\right]^{\frac{3}{2}}}dr
+D_+^0\right\},
\cr
&C_-=C_-^0\e^{B}\sin\left(h
\int^r \frac{\e^{A-B+C}}{\sqrt{\e^{2B}-h^2\e^{2C}}}dr
+D_-^0\right),
\cr
}
\label{gap7}
\end{equation}
where $C_\pm^0$ and $D_\pm^0$ are integration constants.

\

\subsection{Static, axisymmetric spacetime}\label{ap5}

\

A metric of the static, axisymmetric spacetime
is given by
\begin{equation}
ds^2=-\e^{2\alpha}dt^2+\e^{2\beta}dx^2+\e^{2\gamma}dy^2+\e^{2\mu}d\phi^2,
\label{k1}
\end{equation}
where the metric functions, $\alpha$, $\beta$, $\gamma$ and $\mu$
are functions of $x$ and $y$.
A null vector, $k^\mu$, which is tangent to the null geodesic 
on the equatorial plane, $y=0$, is
obtained as 
\begin{equation}
\begin{array}{l}
k^0=\e^{-2\alpha}, \ \ \ k^2=0, \ \ \ k^3=h\e^{-2\mu}, \\
k^1=\pm \e^{-\beta}\sqrt{\e^{-2\alpha}-h^2\e^{-2\mu}},
\end{array}
\end{equation}
where a constant, $h$, is an impact parameter.
The complex null vector, $t^\mu$, becomes
\begin{equation}
\eqalign{
&t^0=\frac{1}{2\sqrt{2}}\left[\frac{2}{h}\e^{\mu}S_{+}S_{-}
+H(S_{+}^2+S_{-}^2)\right]\e^{-\alpha},~~ 
t^2=\frac{i}{\sqrt{2}}\e^{-\gamma}, \cr
&t^1=\frac{1}{2\sqrt{2}}\left[\frac{1}{h}(S_{+}^2+S_{-}^2)\e^{\mu}
+2HS_{+}S_{-}\right]\e^{-\beta},~~
 t^3=\frac{1}{\sqrt{2}}Hh\e^{-2\mu},\cr
&S_{\pm} = \sqrt{\e^{-\alpha}\pm h\e^{-\mu}}, ~~
H=-\frac{1}{h}\int^x
\frac{\e^{\mu-\alpha}}{S_{+}S_{-}}
\frac{\partial\mu}{\partial x}dx.\cr
}
\end{equation}

\




\begin{thebibliography}{99}
\bibitem{1}C. Brans and R. H. Dicke,
          \PR{124,1962,925}. 
\bibitem{2}M. B. Green, J. H. Schwartz and E. Witten,
            {$Superstring$ $Theory$ $vols.$ $1,2$},
         (Cambridge University Press, Cambridge, 1987).
\bibitem{3}T. Damour and G. Esposito-Far\`{e}se,
         Class. Quant. Grav. $\mathbf{9}$, (1992), 2093 .
\bibitem{4}T. Damour and G. Esposito-Far\`{e}se,
          \PRL{70,1993,2220}.
\bibitem{5}T. Damour and G. Esposito-Far\`{e}se,
           \PRD{54,1996,1474} . 
\bibitem{6}T. Tsuchida, G. Kawamura and K. Watanabe, 
             \PTP{100,1998,291}.  
\bibitem{7}T. Tsuchida and K. Watanabe, 
           \PTP{101,1999,73}.  
\bibitem{8}A. Tomimatsu and H. Sato, 
          \PTP{50,1972,95}.
\bibitem{9}B. Voorhees, 
          \PRD{2,1970,2119}.
\bibitem{10}R. K. Sachs, Proc. R. Soc. London, 
            $\mathbf{A264}$, (1961), 309.
\bibitem{11}C. C. Dyer, Mon. Not. R. Astr. Soc., 
            $\mathbf{180}$, (1977), 231. 
\bibitem{14}B. Carter, 
      \JMP{10,1969,70}.
\bibitem{12}R. M. Wald,
          $General$ $Relativity$, 
         (The University of Chicago Press, Chicago and London, 1984).
\bibitem{13}S. Chandrasekhar, 
          $The$ $Mathematical$ $Theory$ $of$ $Black$ $Holes$   
         (Oxford University Press, New York, 1983).
\bibitem{15}$A~passage~from~some~famous~song$.
\end{thebibliography}
\end{document}